\documentclass[cernpreprint, atlasdraft=true, texlive=2016, UKenglish, texmf]{atlasdoc}
 
\usepackage{atlaspackage}
\usepackage{adjustbox}
\usepackage{atlasbiblatex}
 
\usepackage{atlasphysics}
 
\usepackage[output=false]{atlastodo}
 
\graphicspath{{logos/}{figures/}}
 
\usepackage{ANA-HIGG-2018-26-PAPER-defs}

\AtlasTitle{Search for associated production  of a $Z$ boson with an invisibly decaying Higgs boson or dark matter candidates at $\sqrt{s}=13$~\TeV\ with the ATLAS detector}
 
\AtlasAbstract{
A search for invisible decays of the Higgs boson as well as searches for dark matter candidates, produced together with a leptonically decaying $Z$ boson, are presented. The analysis is performed using proton$-$proton collisions at a centre-of-mass energy of 13~\TeV, delivered by the LHC, corresponding to an integrated luminosity of 139~\invfb and recorded by the ATLAS experiment. Assuming Standard Model cross-sections for $ZH$ production, the observed (expected) upper limit on the branching ratio of the Higgs boson to invisible particles is found to be \Hinvobs  (19\%) at the 95\% confidence level. Exclusion limits are also set for simplified dark matter models and two-Higgs-doublet models with an additional pseudoscalar mediator.
}
 
\author{The ATLAS Collaboration}
 
\AtlasRefCode{HIGG-2018-26}
 
\AtlasJournalRef{Phys. Lett. B 829 (2022) 137066}
\AtlasDOI{DOI:10.1016/j.physletb.2022.137066}

\PreprintIdNumber{CERN-EP-2021-204}

\hypersetup{pdftitle={ATLAS document},pdfauthor={The ATLAS Collaboration}}
 
\begin{document}
 
\maketitle
 
\tableofcontents

\section{Introduction}
\label{sec:intro}
% The next lines are included from the .//sections/intro.tex input file
Understanding the nature of dark matter (DM) is one of the most important goals in particle physics today.
Experiments at particle colliders such as the Large Hadron Collider (LHC) might provide sensitivity that complements searches for naturally occurring DM particles or their decay products~\cite{Akerib:2016vxi,Aprile:2018dbl,Agnes:2018ves,Aprile:2019jmx,PandaX-II:2020oim,Akerib:2017kat,Amole:2019fdf,Aprile:2019dbj} by attempting to produce and detect them in the laboratory.
The nature of DM particles remains largely unknown, and there is no obvious candidate in the Standard Model (SM) of particle physics.
One of the most credited hypotheses is that DM candidates are weakly interacting massive particles
(WIMPs, denoted by the symbol $\chi$). At hadron colliders, searches for WIMP-like DM production rely on one or
more visible particles being produced in association with the invisible DM candidates, whose experimental
signature would be the missing transverse momentum (\met) in the collision event.
 
Several models have been proposed in the past decades, with different assumptions about the DM couplings to SM particles and related production processes. As DM particles must be massive, searches for physics beyond the SM in which
DM couples to the Higgs boson are strongly motivated. The Higgs boson was discovered in 2012 by the ATLAS and CMS collaborations at the LHC, with a mass of approximately 125~\GeV~\cite{HIGG-2012-27,CMS-HIG-12-028}.
If the DM particles are in the right mass range, they could even be produced in decays of the Higgs boson. The SM branching ratio prediction for Higgs boson decays into $ZZ \rightarrow 4\nu$ is only 0.1\%~\cite{deFlorian:2016spz}. Assuming SM production of the Higgs boson, its branching ratio to invisible particles, \brHinv, can be constrained in the absence of a significant excess of such events above the expected background.
 
Searches for an excess in events with two electrons or muons\footnote{In this Letter, `lepton' or $\ell$
refers to an electron or muon.} and missing transverse momentum (\MET) are sensitive to Higgs boson decays into DM if the Higgs boson is produced in association with a $Z$ boson that decays into leptons.  The analysis is also sensitive to so-far undiscovered heavier scalars produced together with the $Z$ boson and decaying into invisible particles.
 
Searches in the \Zmet final state can also be used to probe simplified
DM models~\cite{boveia2016recommendations,Albert:2017onk}, in which DM is
produced through a mediator particle that also couples to quarks and
as such searches for dijet resonances are the most
sensitive~\cite{CMS-EXO-16-056,EXOT-2019-03}. In the analysis
presented here, benchmark models are used, and these consider
s-channel production of DM particles through vector or axial-vector
mediators. The models are defined by five parameters: the mediator and
DM particle masses, and the mediator couplings $g_\chi$, $g_q$, and
$g_\ell$ to the DM particles, quarks, and leptons, respectively.
Exclusion limits are set in a plane spanning the DM and mediator
masses, for chosen values of the mediator couplings.
 
\begin{figure}[h!]
\centering
\begin{tabular}{cccc}
\includegraphics[width=0.23\textwidth]{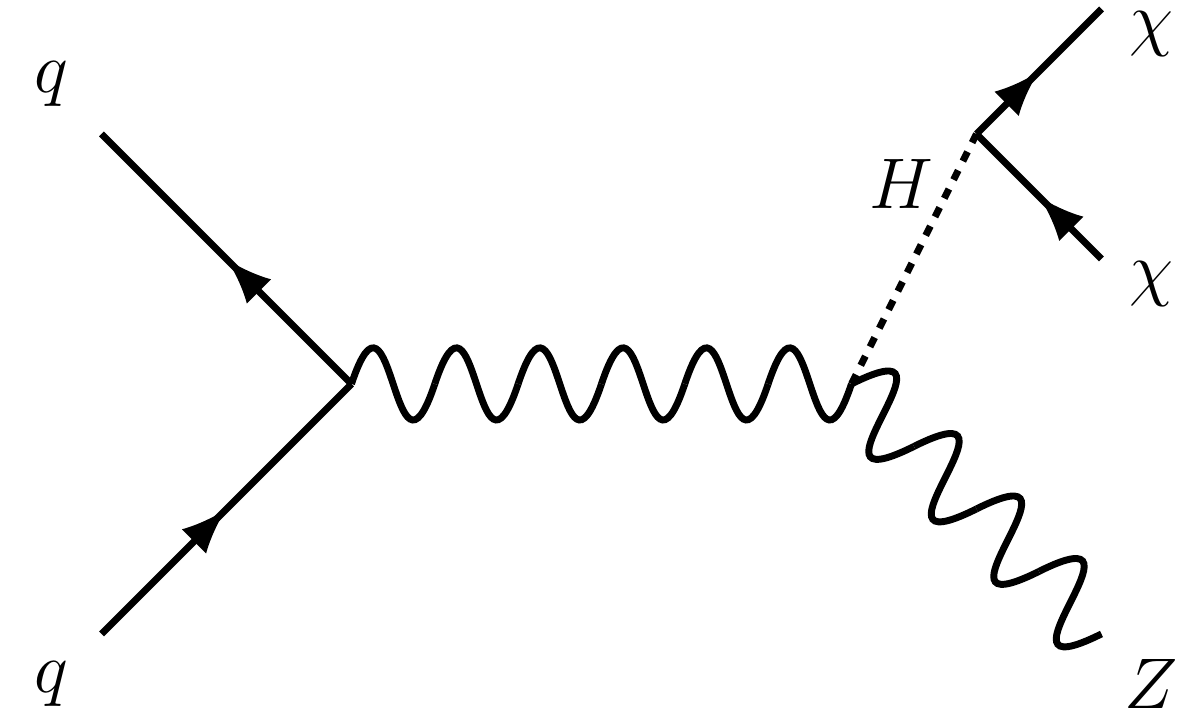} &
\includegraphics[width=0.23\textwidth]{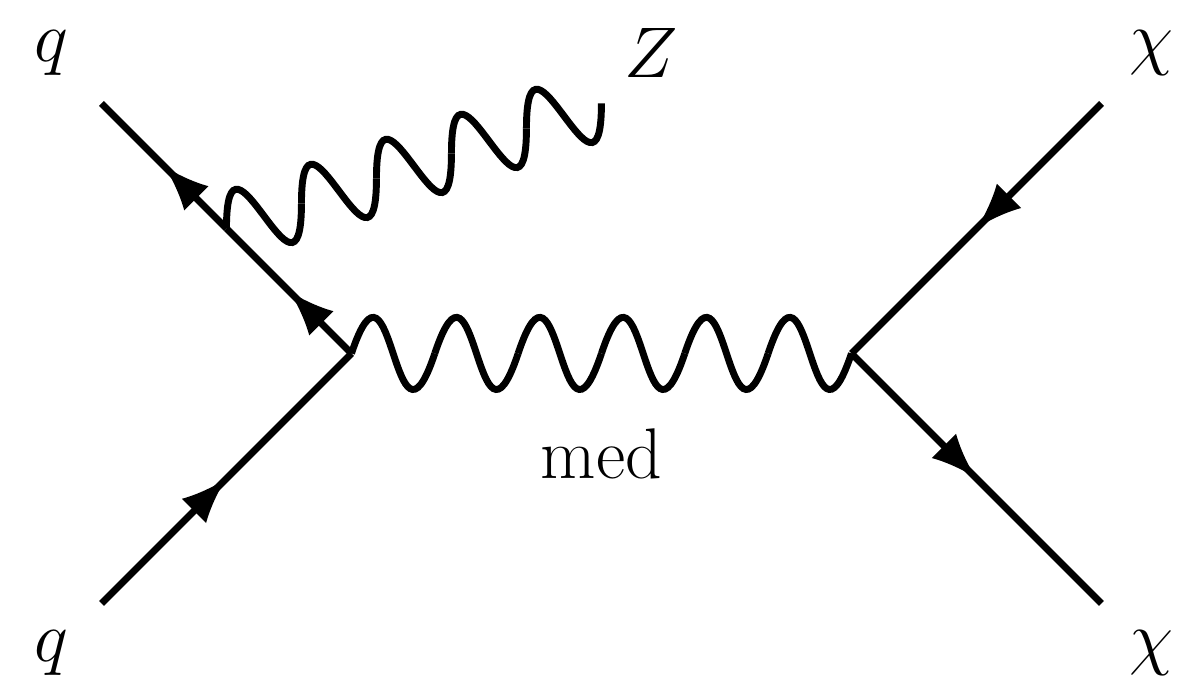} &
\includegraphics[width=0.23\textwidth]{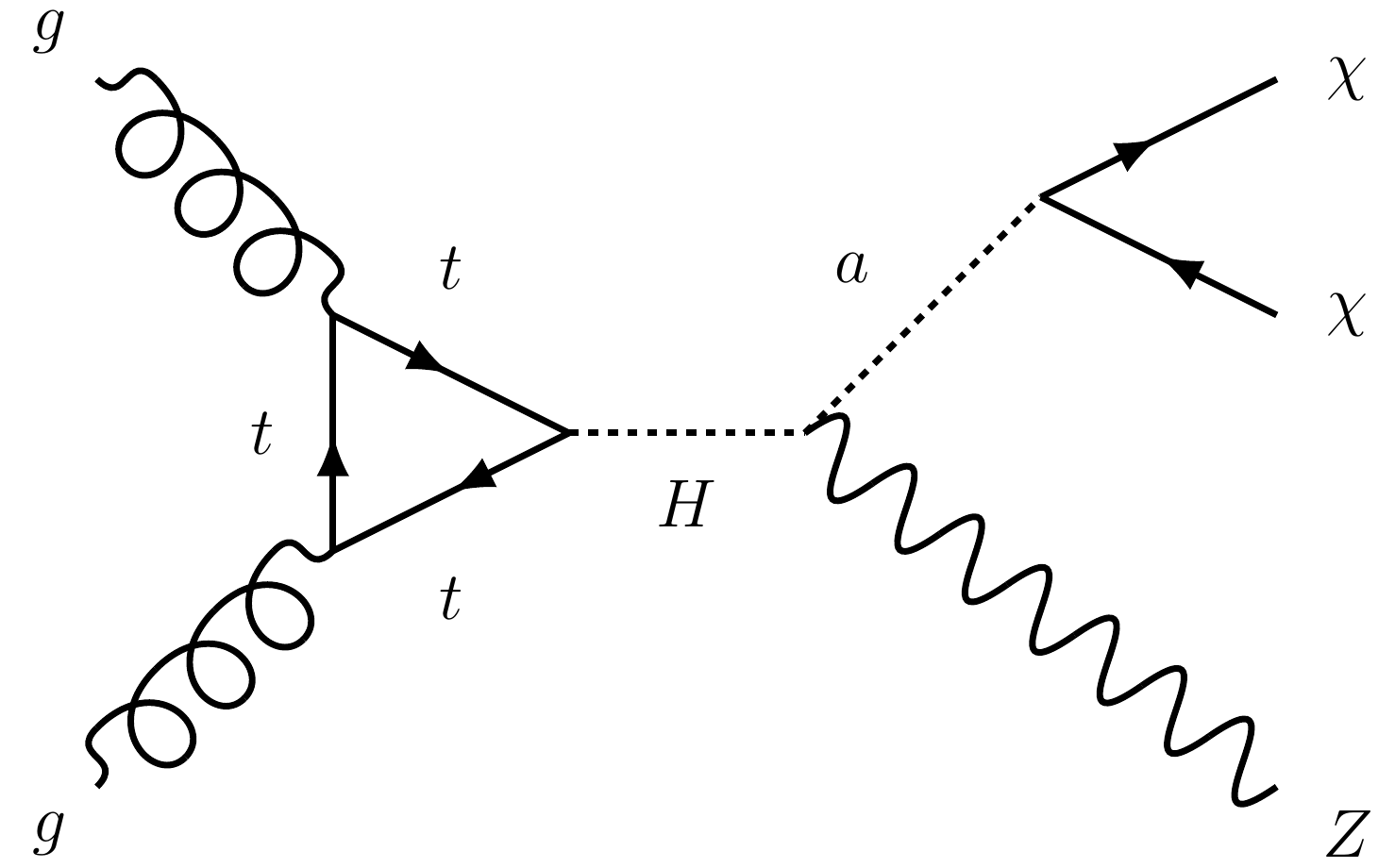} &
\includegraphics[width=0.17\textwidth]{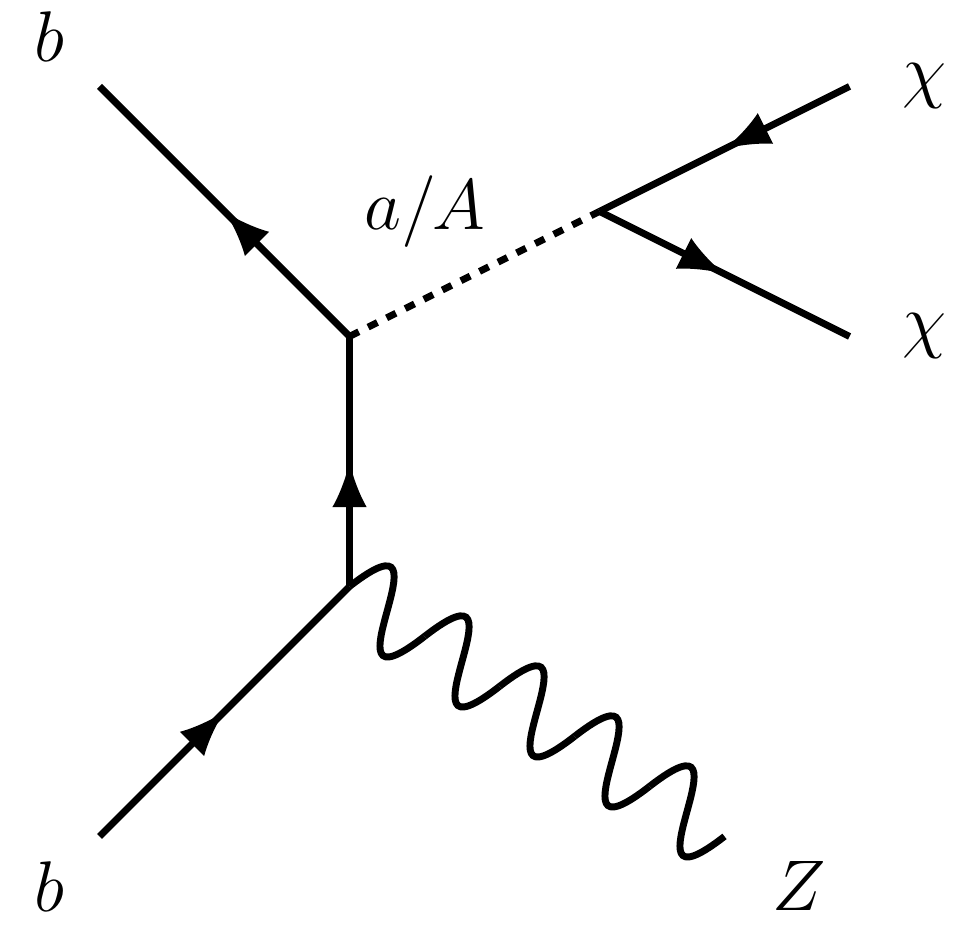} \\
(a) & (b) & (c) & (d) \\
\end{tabular}
\caption{Example Feynman diagrams of the probed processes: (a) associated production of a Higgs boson and a $Z$ boson, where the Higgs boson decays into DM particles, (b) production of a $Z$ boson and a mediator from a quark initial state in the simplified DM models, and (c) $gg$- and  (d) $bb$-initiated \HDMa diagrams.
}
\label{fig:feynman}
\end{figure}

Furthermore, the analysis tests two-Higgs-doublet models (2HDM) that include an additional pseudoscalar mediator $a$ and are called \HDMa~\cite{Bauer:2017ota, Abe:2018bpo, EXOT-2017-32}.
The two Higgs doublets in the model are CP-conserving and of type II~\cite{YR3,Lee:1973iz,Gunion:2002zf,Branco:2011iw}, and the lighter scalar is identified as the observed 125~\GeV\ Higgs boson. Four benchmark scenarios are probed in various planes as a function of the mass of the pseudoscalar Higgs boson, \mA; the mass of the additional pseudoscalar, \ma; the ratio of the vacuum expectation values of the two Higgs doublets, \tanbeta; and \sinTheta, where $\theta$ is the mixing angle between the two CP-odd weak spin-0 eigenstates~\cite{Abe:2018bpo}. Example Feynman diagrams for all probed models are shown in Figure~\ref{fig:feynman}.

The  \Zmet channel is one of the most sensitive for the \Hinv search and the most sensitive channel over much of the parameter space for \HDMa searches.
 
Previous results in the \Zmet final state were obtained with partial Run-2 datasets by the ATLAS Collaboration. Using 36~\invfb of data, ATLAS set a 95\% confidence level (CL) upper limit of 67\% on the Higgs boson branching ratio to invisible particles, with an expected limit of 39\% in the absence of signal~\cite{HIGG-2016-28}. The same paper also reported exclusion limits for the simplified DM models mentioned above. Combining different Higgs boson production channels and different LHC runs, including 36~\invfb from Run~2, ATLAS set an upper limit of 26\% (17\% expected) on the Higgs boson branching ratio to invisible particles~\cite{HIGG-2018-54}.
The CMS Collaboration published results in the \Zmet final state based on 137~\invfb of Run-2 data, including limits on the branching ratio for invisible decays of the Higgs boson (29\% observed vs 25\% expected), simplified DM models and \HDMa models~\cite{CMS-EXO-19-003}. CMS also combined results from different production modes and LHC runs to set an upper limit of 19\% (15\% expected) on the Higgs boson branching ratio to invisible particles~\cite{CMS-HIG-17-023}.
Constraints on \HDMa models were also set by the ATLAS and CMS collaborations using various final states~\cite{EXOT-2017-32,CMS-EXO-16-050,EXOT-2018-43,CMS-EXO-19-003}.

The analysis strategy is outlined briefly in the following. Using the
full LHC Run-2 dataset of 139~\invfb recorded with the ATLAS detector,
events in this search are required to have two oppositely charged
electrons or muons, consistent with originating from a $Z$ boson
decay, as well as significant \MET. The same event selection is
applied in both the \ZHinv search and the other DM searches. A boosted
decision tree (BDT) is trained so that its output is used as the
discriminating observable in the search for invisible Higgs boson
decays, while the searches in the context of simplified DM models and
\HDMa models are based on an observable representing the transverse
mass distribution of the dominant $ZZ$ background. Background distributions
are estimated using simulated samples, and a simultaneous fit is
performed in the signal region and three background control regions to
constrain the systematic uncertainties and determine the normalization
of some of the backgrounds. The sensitivity of the analysis is
considerably improved in comparison with a projection of the previous
analysis scaled to the present integrated luminosity, mainly due to
the use of the BDT and the simultaneous fit of the signal and background control
regions.

% End of text imported from the .//sections/intro.tex input file
 
\section{ATLAS detector}
\label{sec:detector}
% The next lines are included from the .//sections/atlas.tex input file
The ATLAS experiment~\cite{PERF-2007-01} at the LHC is a multipurpose particle detector
with a forward--backward symmetric cylindrical geometry and a near \(4\pi\) coverage in
solid angle.\footnote{ATLAS uses a right-handed coordinate system with its origin at the nominal interaction point (IP)
in the centre of the detector and the \(z\)-axis along the beam pipe.
The \(x\)-axis points from the IP to the centre of the LHC ring,
and the \(y\)-axis points upwards.
Cylindrical coordinates \((r,\phi)\) are used in the transverse plane,
\(\phi\) being the azimuthal angle around the \(z\)-axis.
The pseudorapidity is defined in terms of the polar angle \(\theta\) as \(\eta = -\ln \tan(\theta/2)\).
Angular distance is measured in units of \(\Delta R \equiv \sqrt{(\Delta\eta)^{2} + (\Delta\phi)^{2}}\).}
It consists of an inner tracking detector surrounded by a thin superconducting solenoid
providing a \SI{2}{\tesla} axial magnetic field, electromagnetic and hadronic calorimeters, and a muon spectrometer.
The inner tracking detector covers the pseudorapidity range \(|\eta| < 2.5\).
It consists of silicon pixel, silicon microstrip, and transition radiation tracking detectors.
Lead/liquid-argon (LAr) sampling calorimeters provide electromagnetic (EM) energy measurements
with high granularity.
A steel/scintillator-tile hadronic calorimeter covers the central pseudorapidity range (\(|\eta| < 1.7\)).
The endcap and forward regions are instrumented with LAr calorimeters
for both the EM and hadronic energy measurements up to \(|\eta| = 4.9\).
The muon spectrometer surrounds the calorimeters and is based on
three large superconducting air-core toroidal magnets with eight coils each.
The field integral of the toroids ranges between \num{2.0} and \SI{6.0}{\tesla\metre}
across most of the detector.
The muon spectrometer includes a system of precision chambers for tracking and fast detectors for triggering.
A two-level trigger system is used to select events~\cite{TRIG-2016-01}.
The first-level trigger is implemented in hardware and uses a subset of the detector information
to accept events at a rate below \SI{100}{\kHz}.
This is followed by a software-based trigger that
reduces the accepted event rate to \SI{1}{\kHz} on average
depending on the data-taking conditions.
An extensive software suite~\cite{ATL-SOFT-PUB-2021-001} is used in the reconstruction and analysis of
real and simulated data, in detector operations, and in the trigger and data acquisition systems of the experiment.
 
% End of text imported from the .//sections/atlas.tex input file
 
\section{Data and simulated event samples}
\label{sec:datamcsamples}
% The next lines are included from the .//sections/datamc.tex input file
The presented analysis is performed using data from $pp$ collisions, produced at $\sqrt{s} = 13$~\TeV\ by the LHC and recorded with the ATLAS detector between 2015 and 2018. Data quality requirements~\cite{DAPR-2018-01} are applied to ensure that all detector subsystems were operational. The data correspond to an integrated luminosity of 139~\invfb.
The data sample was collected using a set of
single-electron~\cite{TRIG-2018-05} and single-muon~\cite{TRIG-2018-01}
triggers  which require the presence of an electron (muon)
with  transverse energy \et (transverse momentum \pt) above thresholds in the range of 20--26~\GeV\ depending on the lepton flavour and  data-taking period~\cite{ATLAS:2021lws}. The trigger selections also impose object quality and isolation requirements. There must be a geometrical match between a trigger lepton and a lepton selected in the offline analysis as described in Section~\ref{sec:objselection}.

Simulated Monte Carlo (MC) samples are used to optimize the analysis selection, to estimate the signal, and as input to the background estimation. Unless otherwise mentioned, the \NNPDF[3.0] parton distribution function (PDF) set~\cite{Ball:2014uwa} was used for the hard interaction and the events were passed through the ATLAS detector response simulation~\cite{SOFT-2010-01} within the \GEANT framework~\cite{Agostinelli:2002hh}.
The profiles of the additional inelastic $pp$ interactions (pile-up) in the simulation match those of each dataset
between 2015 and 2018, and were obtained by overlaying minimum-bias events, simulated using the soft QCD processes of
\PYTHIA[8.186]~\cite{Sjostrand:2007gs} with the \NNPDF[2.3lo] PDF set~\cite{Ball:2012cx} and a set of parameters called the A3 tune~\cite{ATL-PHYS-PUB-2016-017}.
For all samples, except those generated with \sherpa~\cite{Bothmann:2019yzt}, the \EVTGEN[1.2.0]~\cite{Lange:2001uf}
program was used to simulate the properties of the $b$- and $c$-hadron decays.

Associated production of a Higgs boson and a $Z$ boson was simulated with \POWHEGBOX[v2]~\cite{Alioli:2010xd}, including both the $qq/qg$ and $gg$ initial states. A Higgs boson mass of 125~\GeV\ was assumed. The \qqZH process was calculated at next-to-leading order (NLO) in QCD, and the \MINLO technique~\cite{Hamilton:2012np} was used to merge 0-jet and 1-jet events. The \ggZH contribution was modelled at leading order (LO) in QCD. The parton-level events were passed to \PYTHIA[8.212]~\cite{Sjostrand:2014zea} with the \AZNLO tune~\cite{STDM-2012-23} to model the Higgs boson decay into four neutrinos as invisible decays, and also the parton showering, hadronization, and multiple parton interactions (MPI). The samples were normalized to next-to-next-to-leading order (NNLO) in QCD with electroweak (EW) corrections (for \qqZH) or to NLO+next-to-leading logarithms (NLO+NLL) in QCD (for \ggZH)~\cite{deFlorian:2016spz,Ciccolini:2003jy,Brein:2003wg,Brein:2011vx,Altenkamp:2012sx,Denner:2014cla,Brein:2012ne,Harlander:2014wda, Harlander:2018yio}. In addition, parameterized EW corrections were applied as a function of the transverse momentum of the $Z$ boson for the \qqZH process~\cite{deFlorian:2016spz}. Allowing the $Z$ boson to decay into electron, muon or $\tau$-lepton pairs, and assuming a Higgs boson branching ratio to invisible particles of 100\%, the production cross-section times branching ratio is $77.0 \pm 1.5$~fb for \qqZH and $12.4 \pm 2.8$~fb for \ggZH~\cite{deFlorian:2016spz}.
 
For the simplified DM models, the s-channel process $pp \rightarrow Z(\ell \ell) \chi \bar{\chi}$ events were simulated with \MGNLO[2.2.2] at NLO in QCD~\cite{Alwall:2014hca}, and fed into \PYTHIA[8.212] with the A14 tune~\cite{ATL-PHYS-PUB-2014-021}. Vector and axial-vector mediator samples were produced for various mediator and DM masses, with the mediator couplings to DM and quarks set to $g_\chi = 1.0$ and $g_q = 0.25$, respectively~\cite{boveia2016recommendations,Albert:2017onk}. Only the leptophobic case is probed, i.e.\ the mediator coupling to leptons was set to $g_\ell = 0$.
These samples were passed through a faster detector simulation using a parameterization of the calorimeter response~\cite{SOFT-2010-01}.
 
\MADGRAPH[5] was used to generate the \HDMa signals, from both the $gg$ and $bb$ initial states, at LO in QCD, in combination with \PYTHIA[8.244] and the A14 tune. The model contains 14 parameters, including the DM mass, the masses of the five Higgs bosons, the ratio of the vacuum expectation values of the two Higgs doublets (\tanbeta), various couplings, and the mixing angles $\alpha$ and $\theta$ between the CP-even and CP-odd weak eigenstates, respectively. The $bb$-initiated production process is particularly important for high values of \tanbeta. Parameters and scanning planes were chosen following the recommendations in Ref.~\cite{Abe:2018bpo}.

Diboson and triboson backgrounds were simulated with \SHERPA[2.2.2], including the parton showering, hadronization and MPI, based on the default \sherpa tunes.
The \ZZllvv and \ZZllll events were simulated for both the  $qq/qg$ and $gg$ initial states and include electrons and muons from $\tau$-lepton decays. The \qqZZ matrix elements were calculated at NLO in QCD for up to one jet, and at LO for two or three jets, with the \comix~\cite{Gleisberg:2008fv} and \OPENLOOPS~\cite{Buccioni:2019sur,Cascioli:2011va,Denner:2016kdg} matrix-element generators. The merging with the \SHERPA parton shower~\cite{Schumann:2007mg} was performed using the \MEPSatNLO prescription~\cite{Hoeche:2012yf}. NLO EW corrections~\cite{Buccioni:2019sur,Kallweit:2014xda,Kallweit:2017khh,Schonherr:2017qcj,Kallweit:2019zez} were applied as a function of \met or the \pt of one of the $Z$ bosons in the 4$\ell$ final state, using the average of additive and multiplicative approaches~\cite{Kallweit:2019zez}. The \ggZZ process was modelled at LO for up to one jet. NLO corrections for  the \ggllvv continuum process are
calculated using the \matrix
package~\cite{Grazzini:2017mhc,Cascioli:2011va,GehrmannManteuffelTancredi2015_1000050906}
with improvements in the calculation of the loop
amplitudes~\cite{Grazzini:2018owa}.
The $WZ$, $WW$ (including \ggWW), and $VVV$ contributions were modelled as well, with the same matrix-element accuracies as the $ZZ$ contributions. The generation includes off-shell effects, $gg\to H$ contributions and interference between processes\cite{Cascioli:2011va}.
 
The \Zjets background was modelled with \SHERPA[2.2.1], with matrix elements calculated at NLO in QCD for up to two jets and at LO for three or four jets.  Top-quark pair ($t\bar{t}$) and single top-quark($Wt$, s-channel and t-channel) production was simulated using \POWHEGBOX[v2] (v1 for t-channel production) at NLO in QCD~\cite{Frixione:2007nw,Nason:2004rx,Frixione:2007vw,Alioli:2010xd,Frederix:2012dh,Re:2010bp}, interfaced to \PYTHIA[8.230] with the A14 tune. 
The predictions were normalised to NNLO+next-to-next-to-leading-logarithm (NNLO+NNLL) cross-section calculations~\cite{Czakon:2011xx, Kidonakis:2011wy, Kidonakis:2010ux,Kidonakis:2010tc}. Associated production of a top-quark pair and a $W$ or $Z$ boson, $t\bar{t}+V$, was simulated with \MGNLO[2.2.3], and interfaced to \PYTHIA[8.210] with the A14 tune. In this case, predictions were normalized to NLO cross-section calculations~\cite{Alwall:2014hca,deFlorian:2016spz}.

% End of text imported from the .//sections/datamc.tex input file
 
\section{Object selection}
\label{sec:objselection}
% The next lines are included from the .//sections/object_selection.tex input file
Selected events are required to contain at least one vertex with a minimum of two associated tracks with $\pT > 500$~\MeV~\cite{ATL-PHYS-PUB-2015-026}. The primary
vertex is chosen to be the vertex reconstructed with the largest
$\Sigma p_{\mathrm{T}}^2$ of its associated tracks.
The event
quality is checked to remove events with noise bursts or coherent
noise in the calorimeters~\cite{DAPR-2018-01}.
 
Electron candidates are reconstructed by matching inner-detector
tracks to clusters of energy deposited in the EM calorimeter.
Electrons must have $\pt>7$~\GeV\ and $|\eta|<2.47$. The
associated track must have $|d_0|/\sigma_{d_0}<5$ and $|z_0|\sin
\theta<0.5$~mm, where $d_0$ ($z_0$) is the transverse (longitudinal)
impact parameter relative to the primary vertex, $\sigma_{d_0}$ is
the uncertainty in $d_0$, and $\theta$ is the polar angle
of the track. Candidates are identified with a
likelihood method and must satisfy `medium' identification
criteria~\cite{EGAM-2018-01}, while `loose' criteria
are used to veto additional electrons in each analysis region.
The likelihood relies
on the shape of the EM shower measured in the calorimeter, the quality
of the track reconstruction, and the quality of the match between the
track and the cluster.
The energy of the EM clusters associated with the electrons is calibrated in successive steps using a combination of simulation-based and data-driven correction factors.  The electron reconstruction, identification, and energy calibration algorithms, as well as their performance, including the associated systematic uncertainties, are studied in Ref.~\cite{EGAM-2018-01}.
Muon candidates are reconstructed in the range $|\eta|<2.5$ by
combining tracks in the inner detector with tracks in the muon spectrometer..
All muon candidates must have $\pt>7$~\GeV,
$|d_0|/\sigma_{d_0}<3$, and $|z_0|\sin \theta<0.5$~mm.
In order to improve the momentum resolution, further quality
requirements are placed on the muons. `Medium' quality
requirements~\cite{PERF-2015-10} are used for candidate muons and `loose' criteria are used to veto additional muons.
The algorithms and efficiency of the muon reconstruction and identification, as well as the momentum calibration, including the associated systematic uncertainties, are estimated as described in Refs.~\cite{PERF-2015-10,MUON-2018-03}.
To suppress hadronic and non-prompt lepton background, electron and muon
candidates are required to satisfy the particle-flow isolation criteria, which are based on tracking and calorimeter measurements~\cite{MUON-2018-03}.

Jets in the range $|\eta|<4.5$ and $\pt>20$~\GeV\ are reconstructed
with a particle-flow algorithm, which combines energy deposits in the
calorimeter with inner-detector tracks~\cite{PERF-2015-09}, using the
anti-$k_t$ algorithm~\cite{Cacciari:2008gp,Fastjet} with a radius
parameter $R$ of $0.4$. A jet-vertex-tagging technique based on a
multivariate likelihood~\cite{PERF-2014-03} is applied to jets with
$|\eta|<2.4$ and $\pt<60$~\GeV\ to suppress jets that are not associated
with the primary vertex of the event. Jets are further calibrated according
to in situ measurements of the jet energy
scale~\cite{JETM-2018-05}. Jets in the range $|\eta|<2.5$ are
identified as  $b$-jets with the MV2c10 algorithm, described in
Ref.~\cite{PERF-2016-05}. The $b$-jet identification efficiency is
about 85\% with a rejection factor of about 25 for light-flavour jets, as measured in
a sample of simulated $t\bar{t}$ events.

Overlaps between reconstructed objects are accounted for with a removal procedure that is mainly
based on the angular separation between the different final-state objects.
The procedure is the same as  the one described in Ref.~\cite{HIGG-2018-09}.

The \metvec of the event is computed as the
negative vectorial sum of the transverse momenta of electrons, muons,
jets  and a track-based soft
term~\cite{PERF-2016-07} that accounts for the contribution from
prompt particles that are not contained in the other objects.
The \MET\ significance~\cite{ATLAS-CONF-2018-038}is defined as $ \metsig = \MET
/(\sigma_\mathrm{L}\sqrt{1-\rho^2_\mathrm{LT}})$, where the parameters
$\sigma_\mathrm{L}$ and $\rho_\mathrm{LT}$ are calculated from MC
simulation and shown to describe the data well; the quantity
$\sigma_\mathrm{L}$ denotes the resolution of the \pt of the system
and $\rho_\mathrm{LT}$ is a correlation factor between resolutions of
the \pt components parallel and perpendicular to the \MET vector.
 
Further related quantities used in the analysis are $H_\mathrm{T}$, the scalar
sum of \pt for all leptons and those jets with $\pt>30$~\GeV, and
$f_\mathrm{ soft}=\left|1 - |\metvec+\sum
\vec{p}_\mathrm{T}^\mathrm{jets}|/\pt^{\ell \ell}\right|$ , a measure
of the fraction of the event's \pt carried by the soft term, where
$\sum \vec{p}_\mathrm{T}^\mathrm{jets}$ is the  vectorial sum of the transverse momenta  of all jets in the
event with $\pt>30$~\GeV\ and $\pt^{\ell \ell}$ is the \pt of the $\ell
\ell$ system.

% End of text imported from the .//sections/object_selection.tex input file
 
\section{Event selection}
\label{sec:evtselection}
% The next lines are included from the .//sections/event_selection.tex input file
Events are selected in a signal region (SR) designed to capture as
many signal events as possible, and three control regions (CR) which
are used to constrain the most important background processes, and which are
expected to have negligible contamination from signal events.
 
Events in the SR are required to have exactly two oppositely charged
electrons or muons with an invariant mass consistent with the mass of the $Z$ boson. The
leptons must have $\pt^\ell>20,30$~\GeV\
when ordered in increasing $\pt$.  The lepton pair is required to have an
invariant mass $m_{\ell \ell}$ in the range $76 < m_{\ell \ell} <
106$~\GeV.  In order to select events in the SR consistent with
invisible particles recoiling against the $Z$ boson, events are
required to have $\MET>90$~\GeV, $\metsig > 9$ and a separation
of $\Delta R_{\ell\ell} <1.8$ between the leptons.
 
An $e\mu$ CR is defined in exactly the same way as the SR apart from
requiring the two leptons to have different flavours.
This CR helps constrain the sum of the non-resonant backgrounds from $t\bar{t}$,
single top-quark, $WW$ and $Z\to\tau\tau$ events. The $e\mu$ CR is about  98\% pure in non-resonant background.
 
A CR containing exactly four leptons, the $4\ell$ CR, is used to constrain
the $ZZ$ background. Its events must contain two pairs of same-flavour
oppositely charged leptons, with the four leptons required to have
$\pt^\ell>7,15,15,27$~\GeV\ when ordered in increasing \pt.  If all four leptons
are of the same flavour, the chosen pairing combination is the one minimizing
the quantity $|m_{\ell \ell 1}-m_Z|+|m_{\ell \ell 2}-m_Z|$, where the
indices $1$ and $2$ denote the lepton pairs and $m_Z$ is taken to be
91.19~\GeV. Both lepton pairs must satisfy $76 < m_{\ell \ell} <
106$~\GeV. In order to mimic the SR, the quantities $\metprime$ and $\metsigprime$ are
calculated in the same way as \MET\ and \metsig, but one pair of leptons, which is
chosen at random, is treated as invisible and excluded from the calculation. The selection
criteria are the same as for the SR, i.e.\ $\metprime>90$~\GeV, $\metsigprime> 9$ and $\Delta
R_{\ell\ell}<1.8$, where the final requirement is imposed on the
remaining lepton pair. The $4\ell$ CR is  almost 100\% pure in $ZZ$ events.
 
A CR containing exactly three leptons, the $3\ell$ CR, is used to
constrain the $WZ$ background. In order
to select the $Z$ boson in the event, two of the leptons must have opposite
charge, the same flavour, $\pt^\ell> 20, 30$~\GeV\ and $76 < m_{\ell \ell} <
106$~\GeV. If there are two combinations that pass these requirements, the one closer in mass to $m_Z$ is taken.
To select events consistent with a $W$ boson decay, it is required that the third lepton
has $\pt^\ell>20$~\GeV,  the event has $\MET>30$~\GeV\ and $\metsig> 3$, and the transverse
mass of the $W$ boson candidate satisfies $m_{\mathrm T}^W=\sqrt{2 \pt^\ell\MET(1-\cos \Delta
\phi(\ell,\met))}>60$~\GeV, where $\Delta \phi(\ell,\met)$ is the
azimuthal angle between the third lepton and $\metvec$. The $3\ell$ CR is  about 93\% pure in $WZ$ events.
 
Events with one or more identified $b$-jets are
removed in all regions to suppress events containing top quarks.
 
Considering all signal events in which the $Z$ boson decays into an electron, muon or $\tau$-lepton pair, the SR selection has an acceptance times
efficiency of $\sim$8\% ($\sim$19\%) for quark-induced (gluon-induced) \ZHinv, resulting in an expectation of 120 events for  $\brHinv = 10\%$.
For an example signal point in the simplified DM model ($\mchi = 1$~\GeV, $\mmed = 900$~\GeV), the acceptance times efficiency is $\sim$20\% with 145 events expected.
For a $gg$-induced \HDMa signal point ($\tanBeta = 1.0$, $\sinTheta = 0.7$, $\mA = 600$~\GeV, $\ma = 400$~\GeV, $\mchi = 10$~\GeV), it is $\sim$32\% with 182 events expected.
For all signal models the acceptance times efficiency is very similar for the dielectron and dimuon selections.
 
% End of text imported from the .//sections/event_selection.tex input file

\section{Background estimates and signal extraction}
\label{sec:sigbkgmodel}
% The next lines are included from the .//sections/sigbkgmodel.tex input file
The signal and all backgrounds are estimated through simultaneous likelihood fits in the signal and control regions, using MC simulation as input. The dominant background in the SR is the $ZZ$ background, followed by
$WZ$, \Zjets, and the non-resonant backgrounds ($WW$, $t\bar{t}$,
single top-quark and $Z\to\tau\tau$), which are treated together. Small
contributions arising from triboson production, $t\bar{t}+V$ and \ZZllll, where two of the
leptons are not reconstructed, are referred to as `Other' in the
following. The $ZZ$ and non-resonant backgrounds have the same
topology as the signal (two leptons and \MET). The SM Higgs boson decay $ZZ\to 4\nu$ is included as part of the signal, while other decays are estimated to be small and neglected.
The $WZ$ background
enters the SR if one of the leptons is not reconstructed. In most of
the \Zjets background, significant \MET arises through mismeasurement
of the energy of the jets or is due to neutrinos or muons that are missed by the reconstruction, coming from
semileptonic heavy-flavour decays.  The normalization of the
MC simulation prediction for \Zjets was
cross-checked using events with $\metsig < 9$, which has a high purity for this process. A sample of $\gamma$ + jets
events, which has similar production diagrams to \Zjets, was used to check
how well the MC simulation described the shape of the data. In both cases the MC simulation prediction and data were observed to agree within the statistical and
systematic errors.

The data are compared with expectation by performing simultaneous
maximum-likelihood fits, using the HistFitter framework~\cite{Baak:2014wma}, to distributions in the signal and control
regions. A separate fit is performed for each signal hypothesis.
Confidence intervals are based on a profile-likelihood-ratio test
statistic~\cite{Cowan:2010js}, assuming asymptotic distributions for
the test statistic.  The CL$_\text{s}$ method~\cite{Cowan:2010js} is used to set exclusion limits. The systematic uncertainties affecting the signal
and background distribution normalizations and shapes across categories are
parameterized by making the likelihood function depend on dedicated
nuisance parameters, constrained by additional Gaussian
probability terms, which are correlated between the regions. The normalizations of the signal, of the $WZ$ background and of the sum of
non-resonant backgrounds are allowed to float in the fits, as their respective CRs have  large numbers of events. The individual components of the non-resonant background
are allowed to vary independently  within their systematic uncertainties.
As the $4\ell$ CR has low statistics the normalization of the $ZZ$ background,
like those of the other backgrounds (\Zjets, triboson production, $t\bar{t}+V$, \ZZllll), can vary only within the systematic uncertainties.

\begin{figure}[!htbp]
\centering
\subfloat[][]{\includegraphics[width=0.45\textwidth]{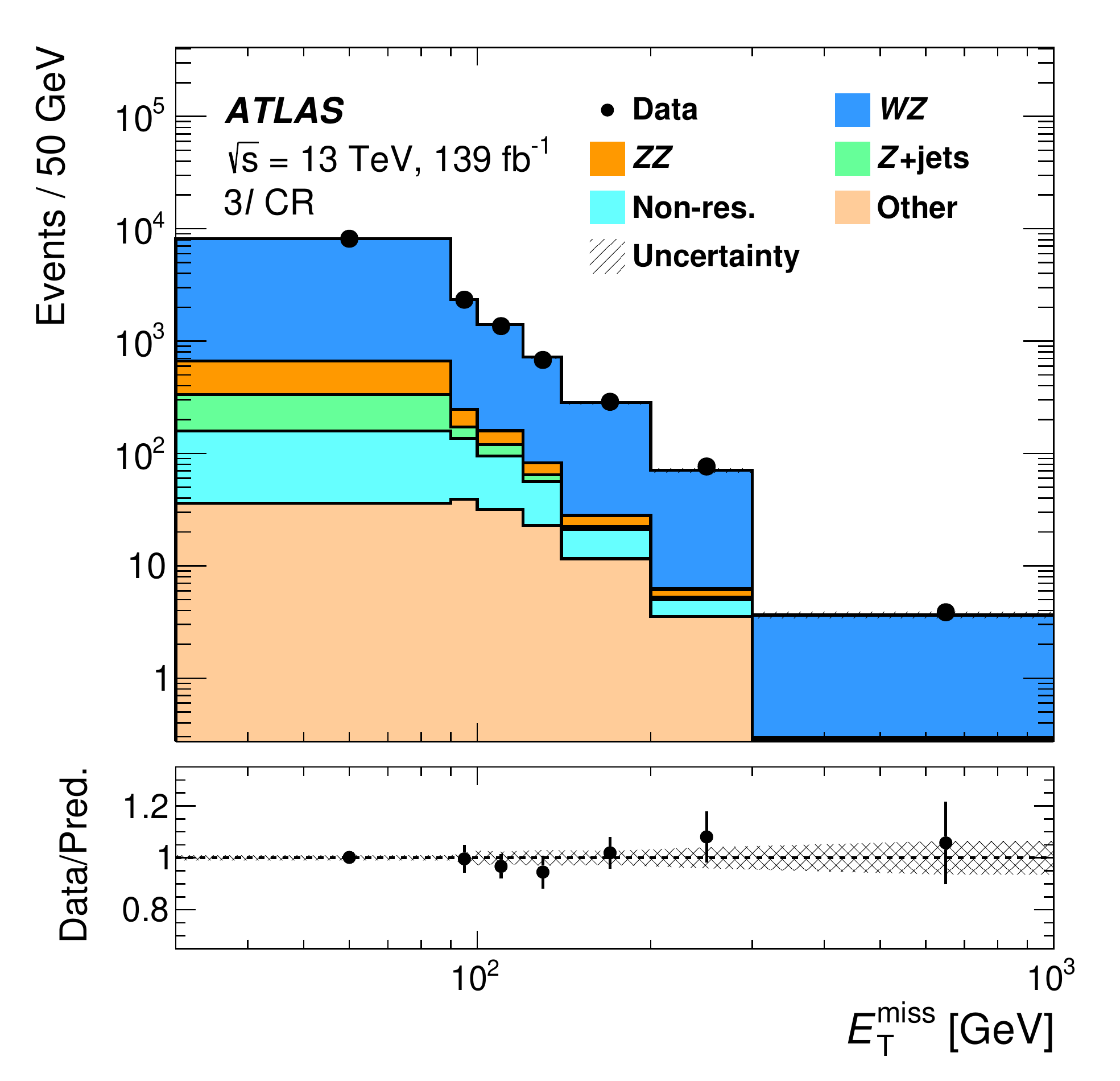}}
\subfloat[][]{\includegraphics[width=0.45\textwidth]{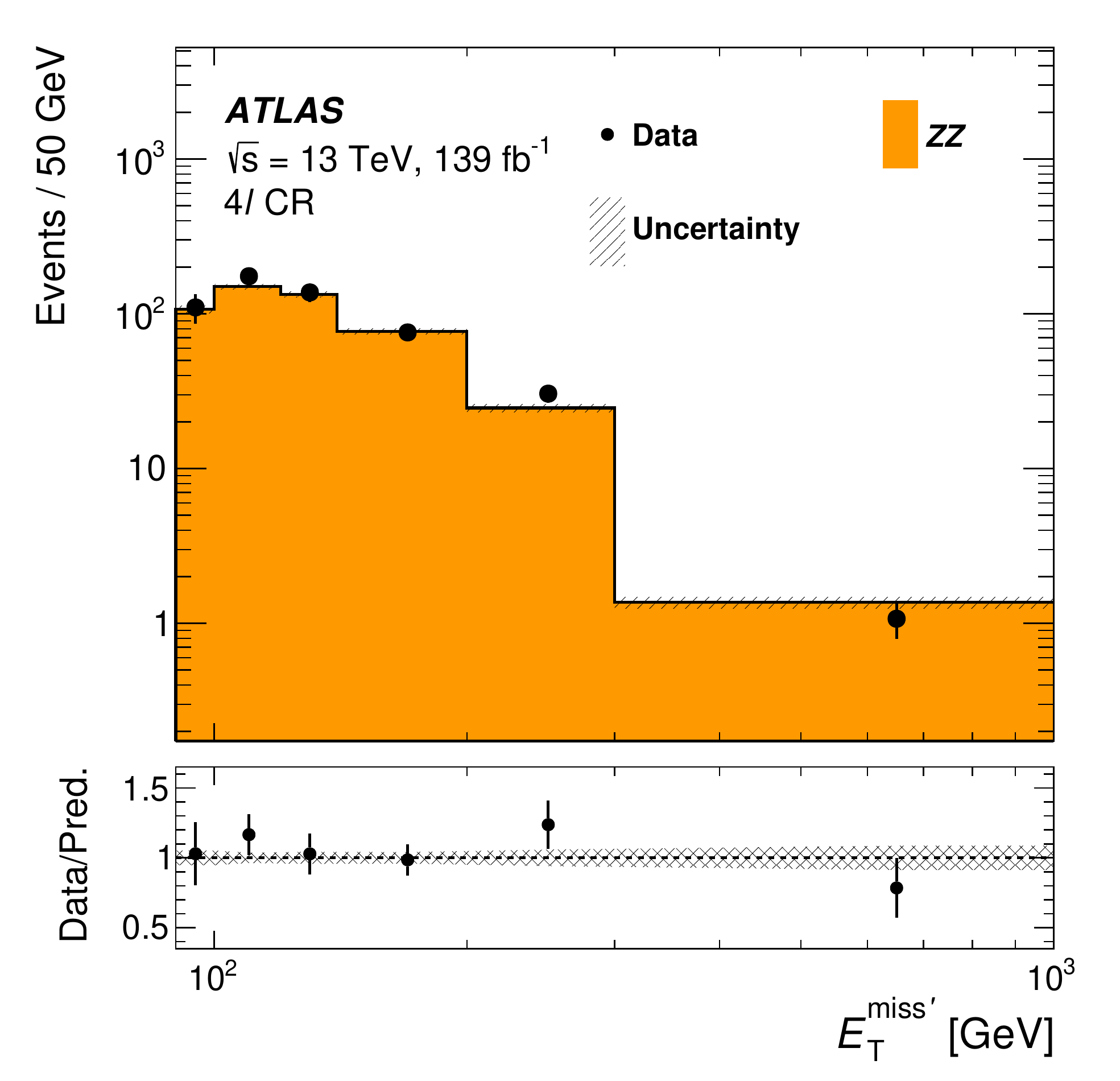}} \\
\subfloat[][]{\includegraphics[width=0.45\textwidth]{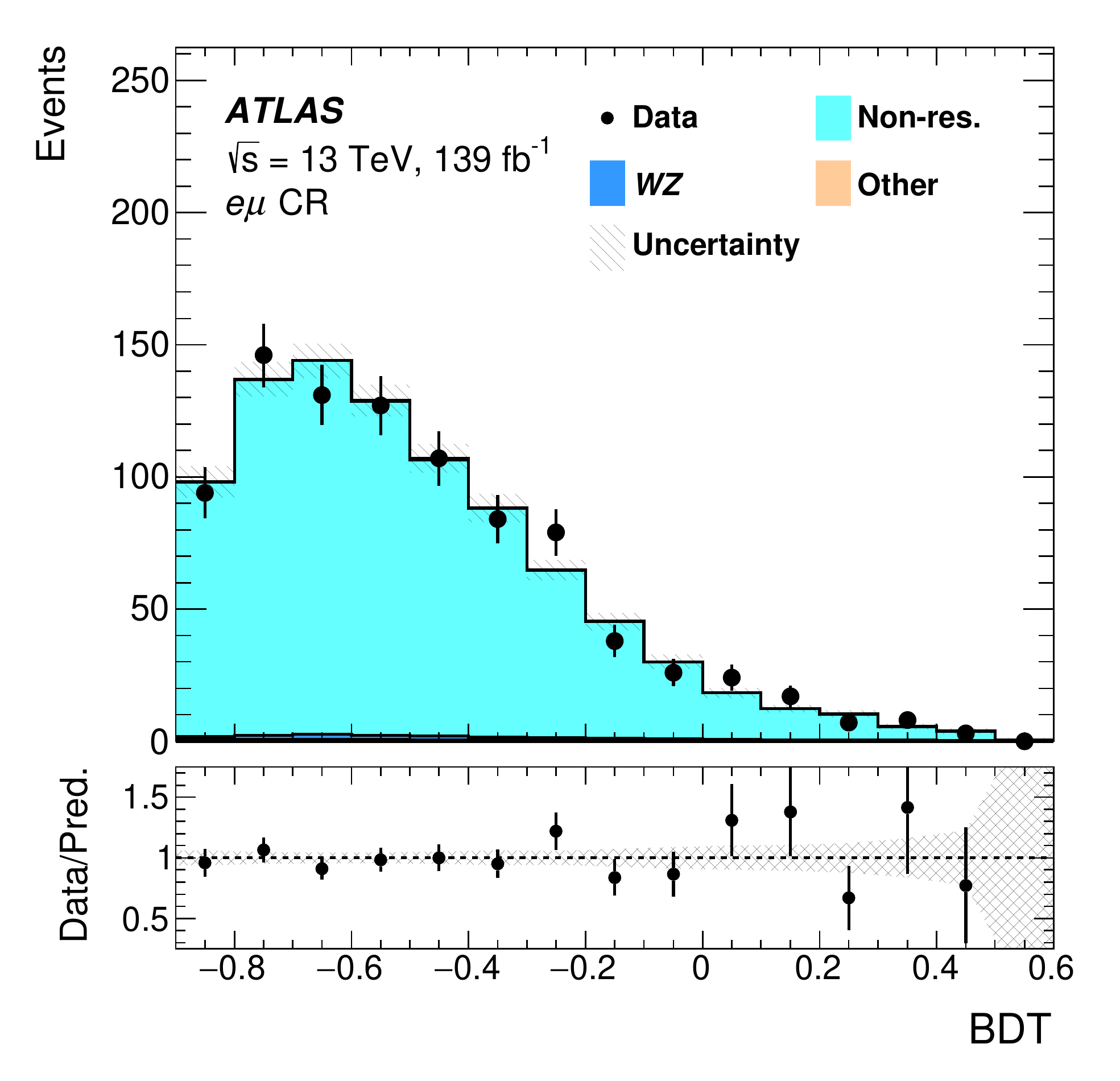}}
\subfloat[][]{\includegraphics[width=0.45\textwidth]{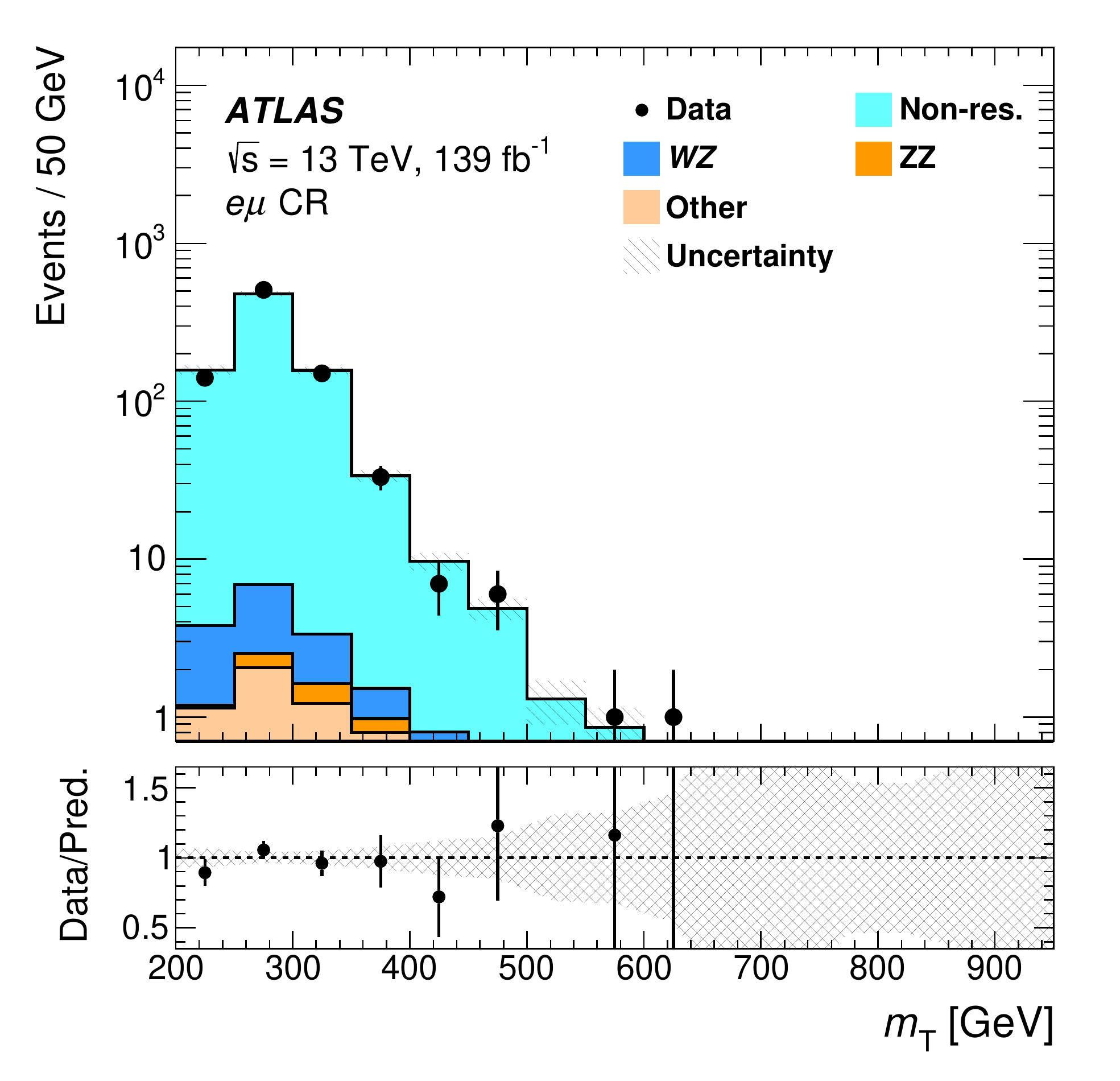}}
\caption{Distributions in data compared with simulated events in the different CRs after (a, b, c) the simultaneous \ZHinv fit and (d) a fit in the context of the simplified DM models (axial-vector signal  with $(\mchi, \mmed) = (150, 900)$~\GeV): (a) \MET distribution in the
$3\ell$ CR,  (b) \metprime distribution in the $4\ell$ CR,  (c) BDT distribution in the $e\mu$ CR,  (d) \mt distribution in the $e\mu$ CR.   As defined in the text, `Non-res.' includes  $WW$, $t\bar{t}$, single-top and $Z\to\tau\tau$ processes, while `Other' stands for triboson, $t\bar{t}+V$, and \ZZllll production. Events recorded below (above) the $x$-range
of the BDT plot are included in the first (last) bin shown. The bottom panel in each figure shows the ratio of the observed data to
the predicted yields. The shaded bands in top and bottom panels represent the total statistical and systematic error of the background.}
\label{fig:CR_postfit_data}
\end{figure}

The sensitivity of the \Hinv search is increased by
using a BDT that is coded in the TMVA
package~\cite{TMVA} to improve the separation between signal and
background events. The BDT uses a set of kinematic distributions that
are selected because they have a different shape for signal and
background. These are combined into a single output variable that
provides better signal/background separation than any of the
individual inputs alone. Eight variables are used: $\MET/H_\mathrm{T}$, \metsig,
$H_\mathrm{T}$, $f_\mathrm{ soft}$, $m_{\ell \ell}$, $\Delta
R_{\ell \ell}$, $y_{\ell \ell}$ (the rapidity of the $\ell \ell$
system), and $\Delta \phi(\ell \ell,\metvec)$ (the azimuthal angle
between the \pT of the $\ell \ell$ system and \metvec). Other variables were
considered but they did not improve the sensitivity significantly.
The BDT was trained in the SR using simulated events for the \ZHinv
signal and for the sum of all backgrounds. The BDT
output distribution is used in the profile likelihood fit for the SR
and $e\mu$ CR. 
The \MET distribution is used for
the $3\ell$ and $4\ell$ CRs as the shapes of many of the BDT input
variables for the background processes differ significantly between the CRs and SR.
Example post-fit distributions in the CRs are shown in
Figures~\ref{fig:CR_postfit_data}(a)--\ref{fig:CR_postfit_data}(c), after the \ZHinv simultaneous fit
to the SR and CRs.  It can be seen that the data are well described by the expectation.
The fit uses the same binning as shown in these plots.

For the searches considering the simplified DM models or \HDMa models,
the transverse mass distribution is used in the maximum-likelihood fits for the SR and the $e\mu$ CR. The quantity
 
\begin{equation}\nonumber
\mt=\sqrt{\left[\sqrt{m_Z^2+(\pt^{\ell\ell})^2}+\sqrt{m_Z^2+(\met)^2}\right]^2-\left[\vec{p}_\mathrm{T}^{\ell\ell}+\metvec\right]^2}
\end{equation}
 
is the transverse mass of the dominant $ZZ$ background,  and gives good separation between signal and background
for the majority of the DM and \HDMa signals considered in this analysis. BDTs were not used for the DM and \HDMa searches, due to the complexity
of training over many mass points. It was found that the \mt distribution  provides better sensitivity over much of the probed parameter space than a BDT trained for one signal point.
The \mt distribution is
only used in the fit for $\mt>200$~\GeV. Figure~\ref{fig:CR_postfit_data}(d) shows that the background estimate is in good agreement with the data for the \mt distribution in the $e\mu$ CR after a simultaneous fit to the signal and control regions in the context of the simplified DM models (axial-vector signal  with $(\mchi, \mmed) = (150, 900)$~\GeV).  As in the \Hinv
search, the \MET distribution is used for the $3\ell$ and $4\ell$ CRs.

% End of text imported from the .//sections/sigbkgmodel.tex input file
 
\section{Systematic uncertainties}
\label{sec:sys}
% The next lines are included from the .//sections/systematics.tex input file
Signal and background expectations are subject to statistical, detector-related and theoretical uncertainties.
The shapes of the observable distributions, the acceptances in signal and control regions, and the overall background sample normalizations that are not free to float in the fit can be affected when varying the simulation to estimate the impact of systematic uncertainties. As discussed in Section~\ref{sec:sigbkgmodel}, the uncertainties are treated as nuisance parameters in the fits and correlated between signal and control regions, which helps to constrain many of them, and reduces their impact. The post-fit impact of the systematic uncertainties can be found in Section~\ref{sec:result}.

Among the systematic uncertainties, the uncertainties in modelling the backgrounds, especially the dominant \ZZllvv background, have the largest impact. Uncertainties in the factorization of NLO EW corrections and higher-order QCD corrections to the \ZZllvv background~\cite{Buccioni:2019sur,Kallweit:2014xda,Kallweit:2017khh,Schonherr:2017qcj} are taken from the deviation of the additive and multiplicative approaches from their average~\cite{Kallweit:2019zez}.
For all diboson backgrounds, uncertainties due to missing higher orders in the QCD calculation are estimated by varying the renormalization and factorization scales by a factor of two, either
independently or correlated and taking the largest variation as the
uncertainty, while the effects of PDF and \alphas uncertainties are
calculated using the PDF4LHC
prescription~\cite{Butterworth:2015oua}. The impact of the choice of
parton shower and hadronization model is evaluated by comparing the samples from the nominal generator set-up with samples produced with varied resummation and CKKW matching procedure~\cite{Catani:2001cc,Hoeche:2009rj}, and a different recoil scheme tuned in \SHERPA.
 
The effects of QCD uncertainties on the \Zjets background are also non-negligible. They are estimated by varying the renormalization and factorization scales by a factor of two. PDF uncertainties are also included. Modelling uncertainties in the \ttbar and single-top-quark backgrounds are subdominant; their estimation is performed by varying the generator used for the hard scatter, as well as the renormalization and factorization scales, the parton shower model, \alphas, and the PDF sets. The modelling uncertainties of the remaining background processes are also evaluated but have no impact on the final result.

For the \ZHinv signal, the uncertainty due to missing higher orders in QCD is estimated in bins of the kinematic observables using a scheme similar to the one discussed in Ref.~\cite{HIGG-2016-21}. The PDF uncertainties are evaluated using the PDF4LHC prescription~\cite{Butterworth:2015oua}. To estimate the impact of using a different parton shower model, the default \PYTHIA[8] parton shower algorithm is replaced with \HERWIG[7]~\cite{Bellm:2015jjp}. Uncertainties due to the EW corrections are estimated by taking the difference between the additive and multiplicative
corrections.
The effects of QCD scale uncertainties for the simplified DM model and the \HDMa prediction are evaluated by varying the scales as described for the diboson backgrounds. The parton shower uncertainty is estimated with the eigenvariations of the A14 tune. The PDF uncertainty includes variations of the \NNPDF[3.0] set, as well as a comparison with two alternative PDF sets.
 
The detector-related uncertainties affecting signal and background predictions are dominated by jet reconstruction uncertainties.
Uncertainties in the jet energy scale (JES)~\cite{JETM-2018-05} arise from the calibration and are derived as function of the jet \pt and  $\eta$. Further contributions emerge from the jet flavour composition and the pile-up conditions.
Uncertainties in the jet energy resolution (JER) depend on the jet \pt\ and $\eta$ and arise both from the method used to derive the resolution and from the difference between simulation and data~\cite{JETM-2018-05}.
Uncertainties in the lepton identification and lepton energy/momentum scale and resolution are included in the fits. These are derived using simulated and measured events with \Zll, \Jpsill and \Wlnu decays~\cite{EGAM-2018-01,MUON-2018-03}.
The uncertainties in the lepton and jet energy scales are propagated to the uncertainty in the \MET~\cite{PERF-2016-07}.
Additionally, the uncertainties from the momentum scale and resolution of the
track-based soft term are included.
The uncertainty in the combined 2015--2018 integrated luminosity is 1.7\%~\cite{ATLAS-CONF-2019-021}, obtained using the LUCID-2 detector~\cite{LUCID2}.

% End of text imported from the .//sections/systematics.tex input file
 
\section{Results}
\label{sec:result}
\begin{figure}[!htbp]
\centering
\subfloat[][]{\includegraphics[width=0.45\textwidth]{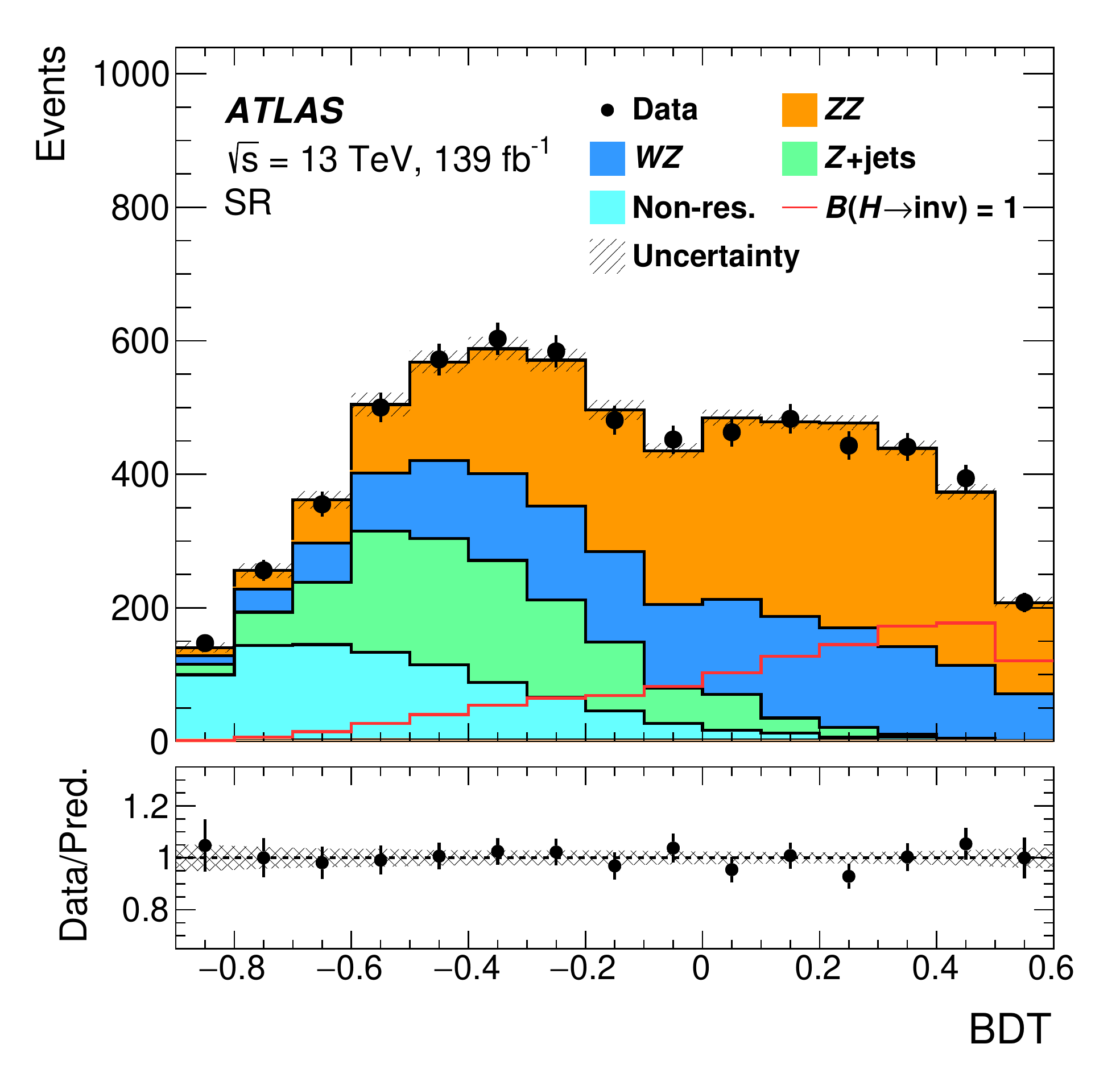}}
\subfloat[][]{\includegraphics[width=0.45\textwidth]{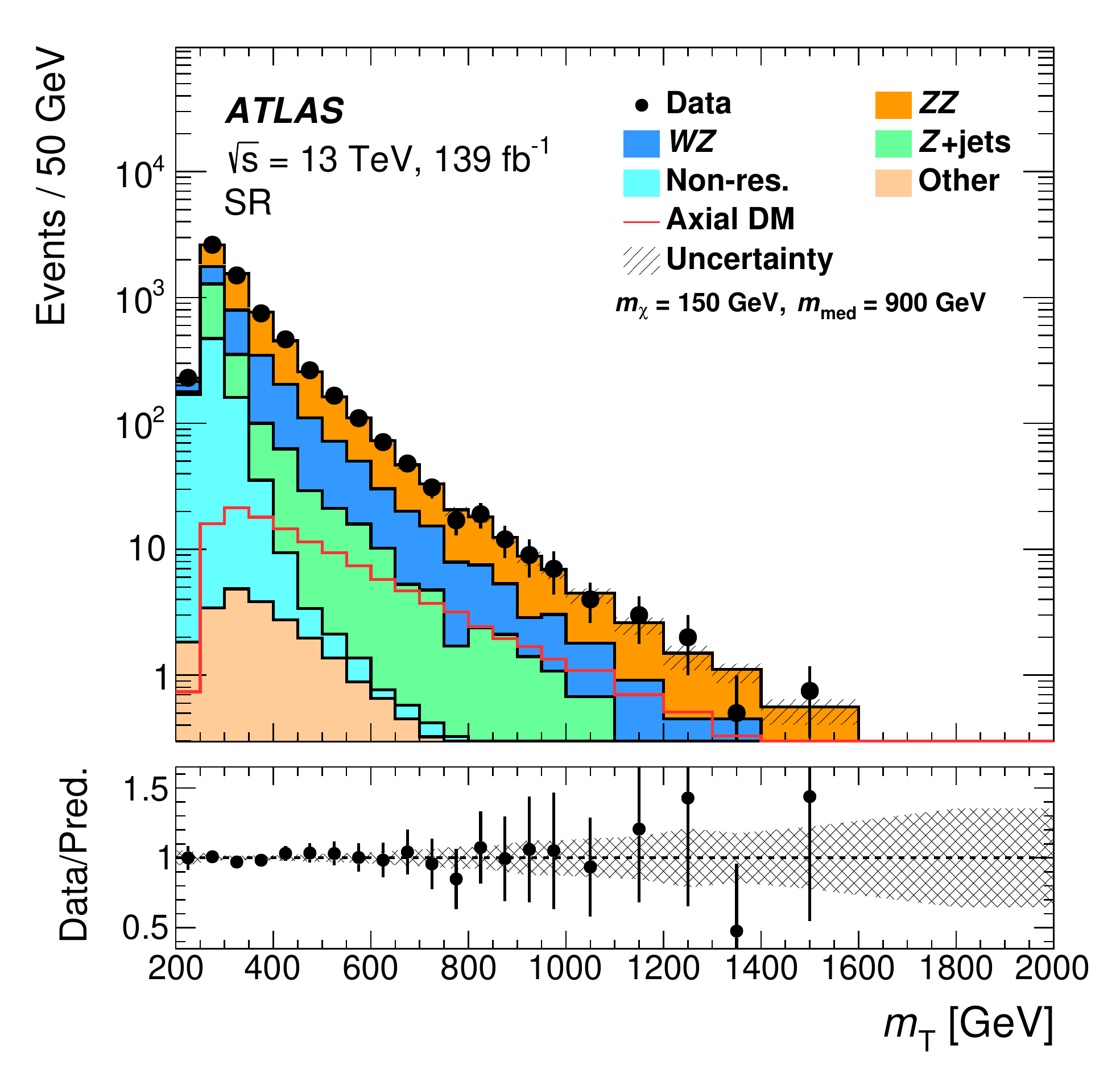}}
\caption{ (a) BDT and (b) \mt distributions in the signal
region, after the respective simultaneous fits. The
\Hinv signal,  with a branching ratio of $\br = 100\%$, is shown in (a)
and an example DM axial-vector signal  with $(\mchi, \mmed) = (150, 900)$~\GeV\ is shown in (b). As defined in the text, `Non-res.' includes  $WW$, $t\bar{t}$, single top-quark and $Z\to\tau\tau$ processes, while `Other' stands for triboson, $t\bar{t}+V$, and \ZZllll  production. Events recorded below (above) the $x$-range
of the BDT plot are included in the first (last) bin shown. The bottom panel in each figure shows the ratio of the observed data to
the predicted yields. The shaded bands in top and bottom panels represent the total statistical and systematic error of the background.}
\label{fig:SR_postfit_data}
\end{figure}

% The next lines are included from the .//sections/results.tex input file
Figure~\ref{fig:SR_postfit_data} shows the BDT and \mt distributions of the selected data events, compared with the estimated backgrounds after the simultaneous fits described in Section~\ref{sec:sigbkgmodel} are performed. For the \ZHinv case, Table~\ref{table:Htoinv_yield_table} gives the number of observed events as well as the signal and background expectations after the simultaneous fit. The numbers for the DM fits are very similar, with some differences in the $e\mu$ CR as the \mt distribution is
only used in the fit for $\mt>200$~\GeV.
 
\newenvironment{DIFnomarkup}{}{}
\begin{DIFnomarkup}

\begin{table}[htbp]
\begin{center}
\caption{\label{table:Htoinv_yield_table}Summary of signal and control region yields after the simultaneous fit for the \ZHinv signal. Also given is the total post-fit uncertainty of each number. As defined in the text, `Non-resonant' includes  $WW$, $t\bar{t}$, single top-quark and $Z\to\tau\tau$ processes.  Note the uncertainty on the total expectation does not equal
the sum of the uncertainties of individual contributions added in quadrature, due to correlations between the uncertainties.}
\setlength{\tabcolsep}{0.0pc}{
\begin{tabular*}{\textwidth}{@{\extracolsep{\fill}}lcccc}
\hline \hline
& {SR} & {$e\mu$ CR} & {3$\ell$ CR} & {4$\ell$ CR}\\[-0.05cm]
\hline
Observed events          & $6382$ & $891$ & $11622$ & $314$ \\
\hline
Expected yields after fit         & $6385\pm81$~~         & $895\pm29$~~           & $11620\pm110$~~           & $296\pm11$~~         \\
\hline
\ZHinv         & ~~~~~~$4 \pm 110$          & -          & ~~-          & -              \\
\ZZllvv         & $2681 \pm 110$          & $0.763 \pm 0.064$          & ~~$2.61 \pm 0.18$          & -              \\
$WZ$         & $1595 \pm 34$~~          & $11.6 \pm 1.1$~~          & $10623 \pm 150$~~          & -              \\
\Zjets         & $1111 \pm 100$          & $0.79 \pm 0.30$          & $235 \pm 89$          & -              \\
Non-resonant         & $881 \pm 39$          & $876 \pm 29$~~          & $220 \pm 31$          & -              \\
\ZZllll         & $85.8 \pm 5.5$          & $0.621 \pm 0.056$          & $443 \pm 40$           & $295 \pm 11$~~              \\
$t\bar{t}+V$         & $12.7 \pm 2.8$          & $1.76 \pm 0.41$          & ~~$53 \pm 12$           & -              \\
Triboson         & $13.0 \pm 6.2$          & $3.1 \pm 1.4$          & ~~$44 \pm 20$          & $0.48 \pm 0.23$             \\
 
\hline \hline
\end{tabular*}
}
\end{center}
\end{table}

\end{DIFnomarkup}

The data are in good agreement with the background expectation.
Assuming SM Higgs production cross-sections and their uncertainties,
the best-fit branching ratio of Higgs boson decays into invisible
particles is $\brHinv = (0.3 \pm 9.0)\%$, which corresponds to an
observed 95\% CL upper limit of \Hinvobs. The corresponding expected upper limit
is \Hinvexp. The analysis improvements and reduced systematic
uncertainties lead to this limit being 45\% lower than a projection of the previous
expected limit~\cite{HIGG-2016-28} scaled to the present integrated luminosity.
The observed normalization factors for the $WZ$ and $e\mu$
backgrounds are \HinvmuWZ and \Hinvmuemu respectively.
Table~\ref{tab:sys:BR} shows the contributions of the statistical and
systematic uncertainties to the total uncertainty of the best-fit
\brHinv.  As discussed in Section~\ref{sec:sys}, the dominant uncertainty
is due to the $ZZ$ modelling. Uncertainties in the jet reconstruction
are important as well.
 
\begin{DIFnomarkup}
\begin{table}[htbp]
\centering
\caption{\label{tab:sys:BR} Summary of the uncertainties  $\Delta\br$  on the best-fit \brHinv, obtained by fixing the corresponding nuisance parameters to their best-fit values, and subtracting the square of the resulting uncertainty from the square of the total uncertainty to evaluate $(\Delta\br)^2$. The statistical uncertainty component is obtained by fixing all nuisance parameters to their best-fit values. Note the total uncertainty does not equal the sum of the individual contributions added in quadrature due to correlations between the systematic uncertainties.}
\renewcommand{\arraystretch}{1.3}
\begin{tabular}{lr}
\hline\hline
Uncertainty source                               & $\Delta\br$ [\%] \\
\hline
Statistical uncertainty                          & 5.1 \\
\hline
Systematic uncertainties                         & 7.4 \\
~~~~~Theory uncertainties                        & 4.9 \\
~~~~~~~~~~Signal modelling                       & 0.4 \\
~~~~~~~~~~$ZZ$ modelling                         & 4.4 \\
~~~~~~~~~~Non-$ZZ$ background modelling          & 2.1 \\
~~~~~Experimental uncertainties (excl.\ MC stat.) & 4.6 \\
~~~~~~~~~~Luminosity, pile-up                     & 1.5 \\
~~~~~~~~~~Jets, \MET                             & 4.0 \\
~~~~~~~~~~Flavour tagging                        & 0.4 \\
~~~~~~~~~~Electrons, muons                       & 1.2 \\
~~~~~MC statistical uncertainty                  & 1.6 \\
\hline
Total uncertainty                                & 9.0 \\
\hline\hline
\end{tabular}
\end{table}
 
\end{DIFnomarkup}
 
\FloatBarrier

\begin{figure}[!htbp]
\centering
\subfloat[][]{\includegraphics[width= 0.49\textwidth]{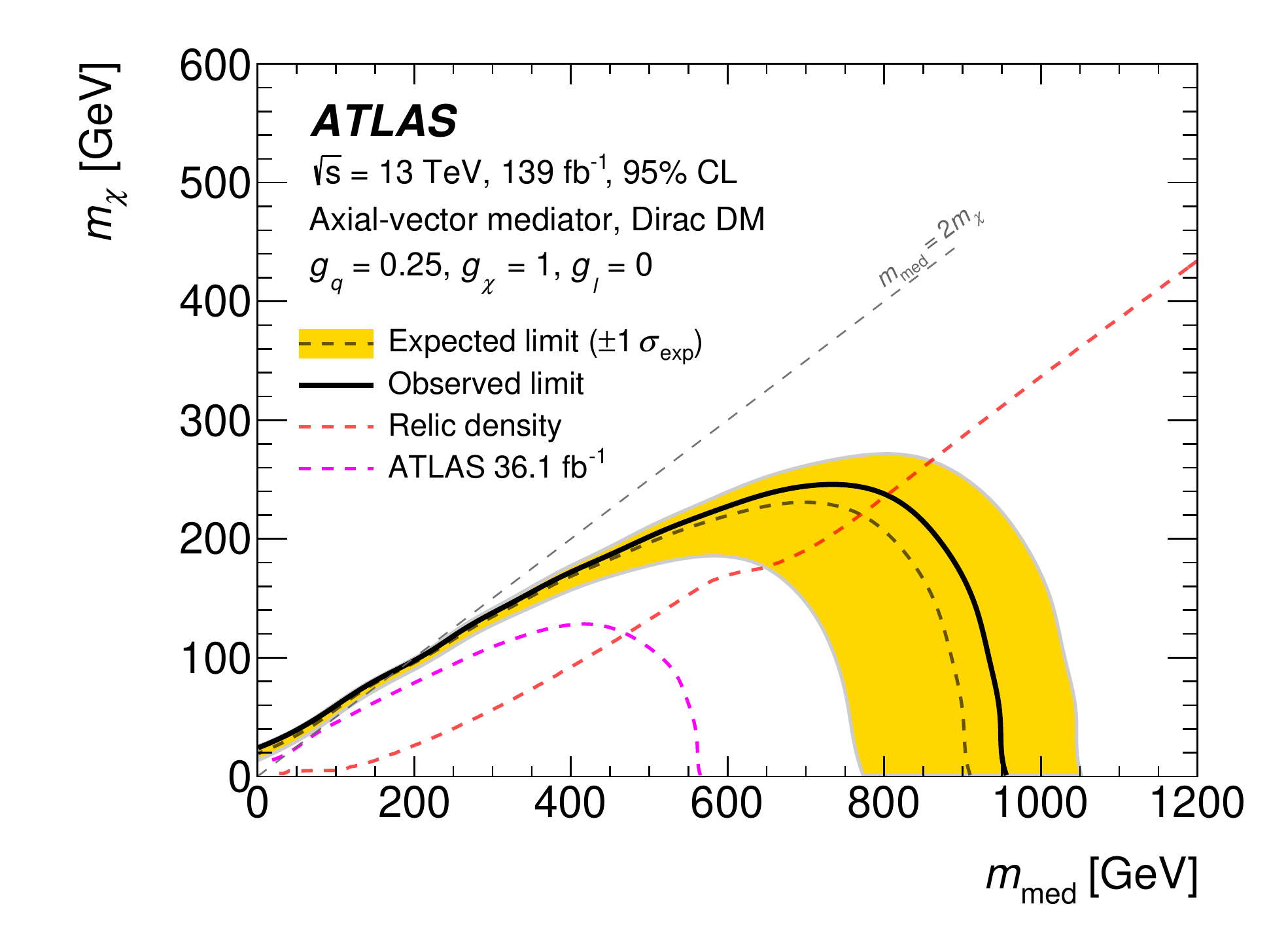}}
\subfloat[][]{\includegraphics[width= 0.49\textwidth]{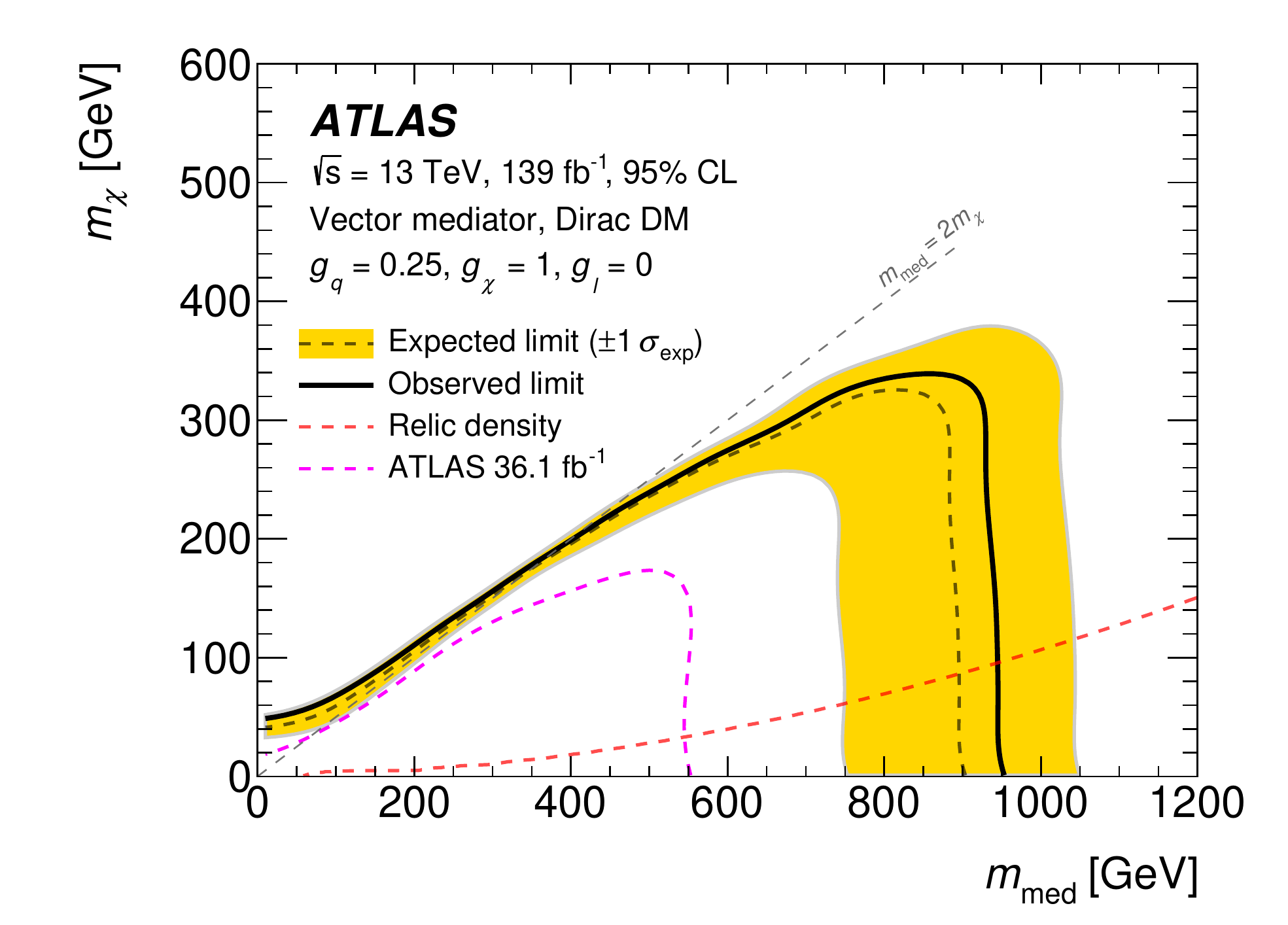}}
\caption{Exclusion limits for simplified DM models with $g_\chi = 1.0$, $g_q = 0.25$, and $g_\ell = 0$, when assuming (a) an axial-vector mediator or (b) a vector mediator. The region below the solid black line is excluded at the 95\% CL. The dashed black line indicates the expected limit in the absence of signal, and the yellow band the corresponding $\pm 1 \sigma$ uncertainty band.
The dashed red line labelled `Relic density' corresponds to combinations of DM and mediator mass values that are consistent with a DM density of $\Omega h^2 = 0.118$ and a standard thermal history, as computed in Ref.~\cite{Albert:2017onk}.
Below the line, annihilation processes described by the simplified model mostly predict too high a relic density while regions with too low a relic density are mostly found for \mmed closer to the DM mass.  The dashed magenta line indicates the previous ATLAS result from a 36.1~\invfb dataset~\cite{HIGG-2016-28}. }
\label{fig:simplified_contours}
\end{figure}

Figure~\ref{fig:simplified_contours} shows the exclusion limits for the simplified DM models assuming either an axial-vector or vector mediator, with the chosen mediator couplings $g_\chi = 1.0$, $g_q = 0.25$, and $g_\ell = 0$.
The bottom-left region inside the solid black contour is excluded at the 95\% CL.
The dashed red line labelled `Relic density' corresponds to combinations of DM and mediator mass values that are consistent with a DM density of $\Omega h^2 = 0.118$ and a standard thermal history, as computed in Ref.~\cite{Albert:2017onk}.
The dashed magenta line indicates the previous ATLAS result from a 36.1~fb$^{-1}$ dataset~\cite{HIGG-2016-28}. The present analysis enlarges the excluded area significantly.  Axial-vector and vector mediators with masses of up to 950~\GeV\ are excluded now, while DM masses of up to 250~\GeV\ are excluded in the axial-vector case and up to 350~\GeV\ in the case of a vector mediator.

\begin{figure}[!htb]
\begin{center}
\begin{tabular}{cc}
\includegraphics[width= 0.45\textwidth]{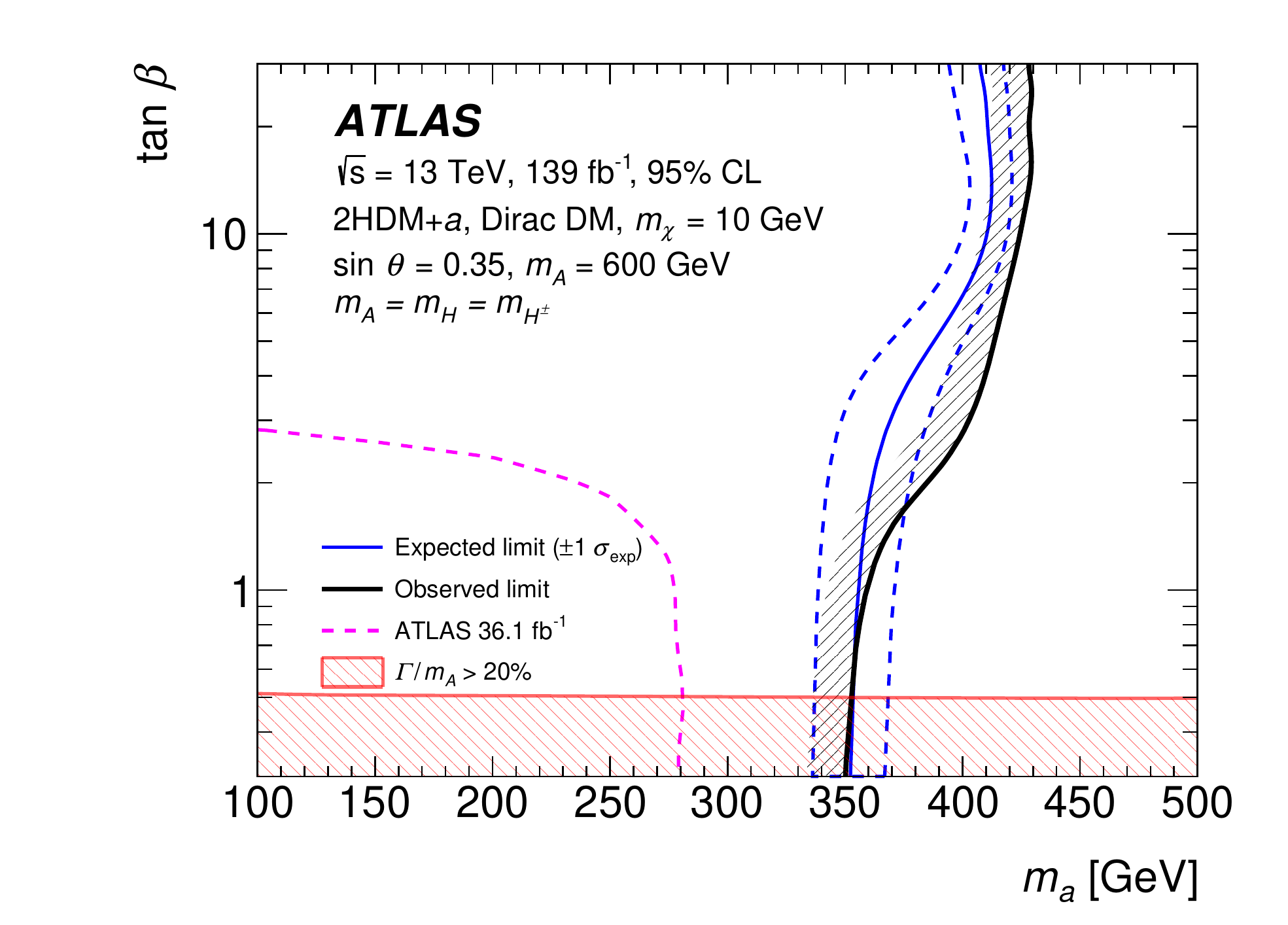} &
\includegraphics[width= 0.45\textwidth]{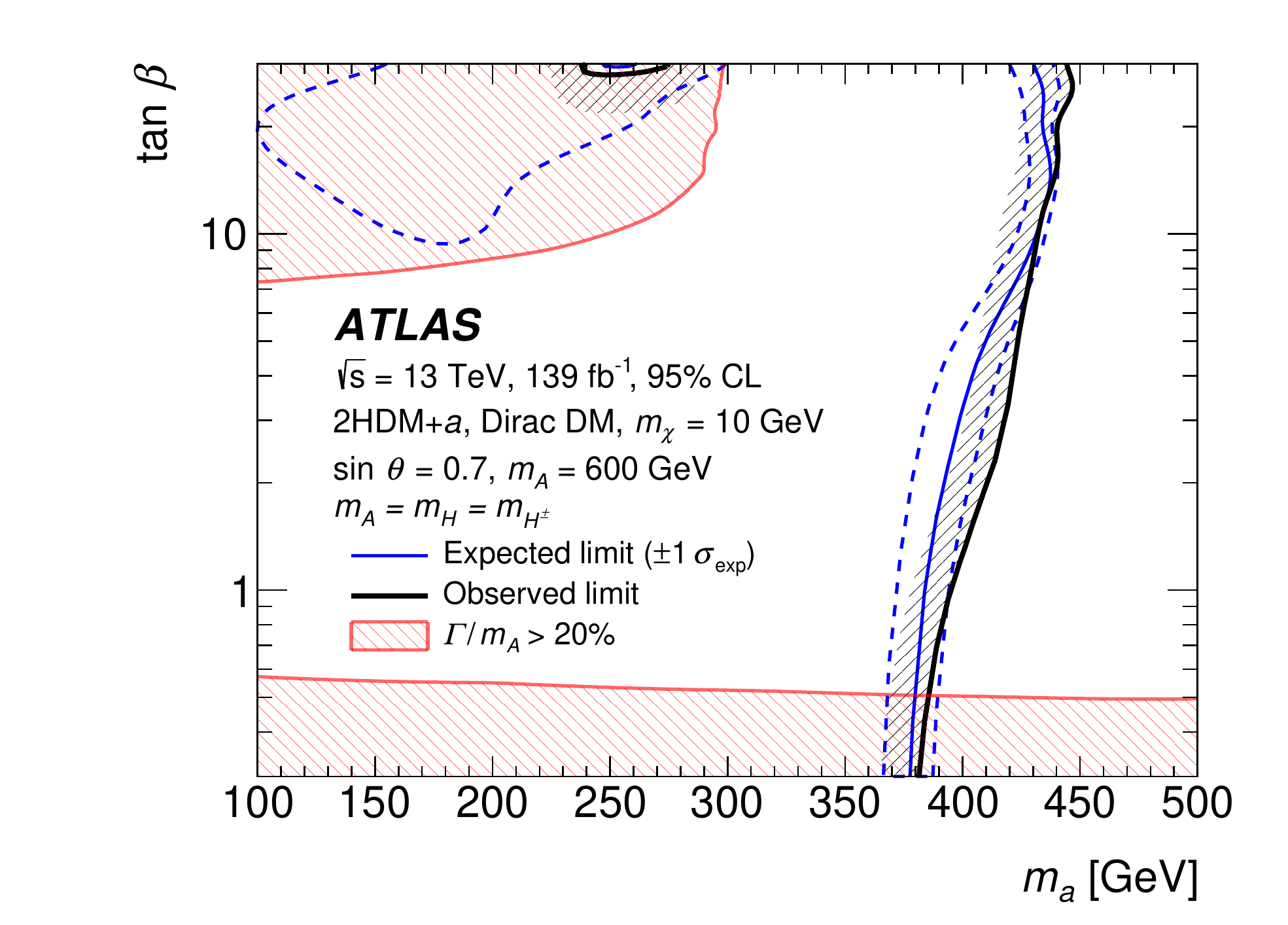} \\
(a) & (b) \\
\includegraphics[width= 0.45\textwidth]{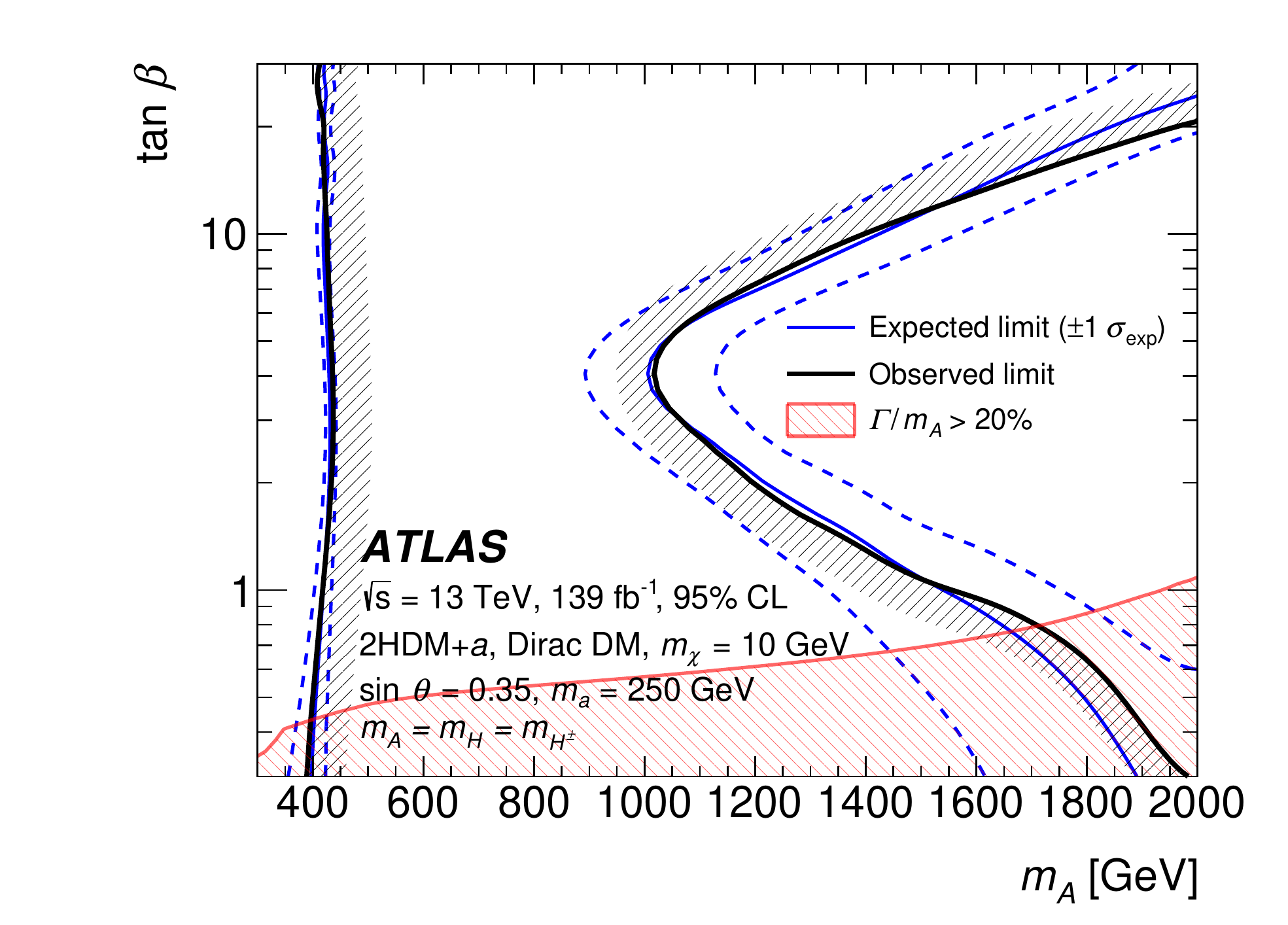} &
\includegraphics[width= 0.45\textwidth]{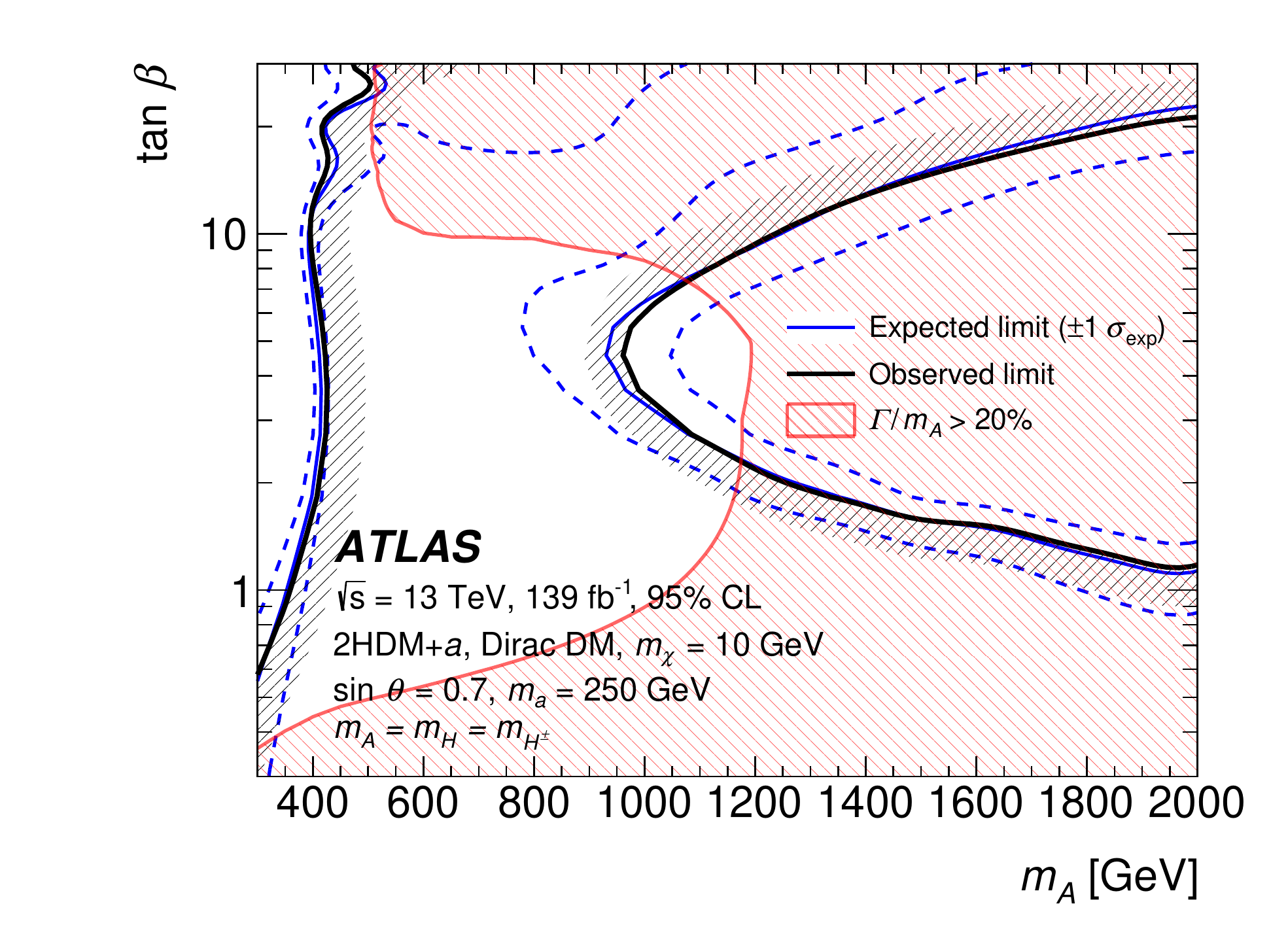} \\
(c) & (d) \\
\includegraphics[width= 0.45\textwidth]{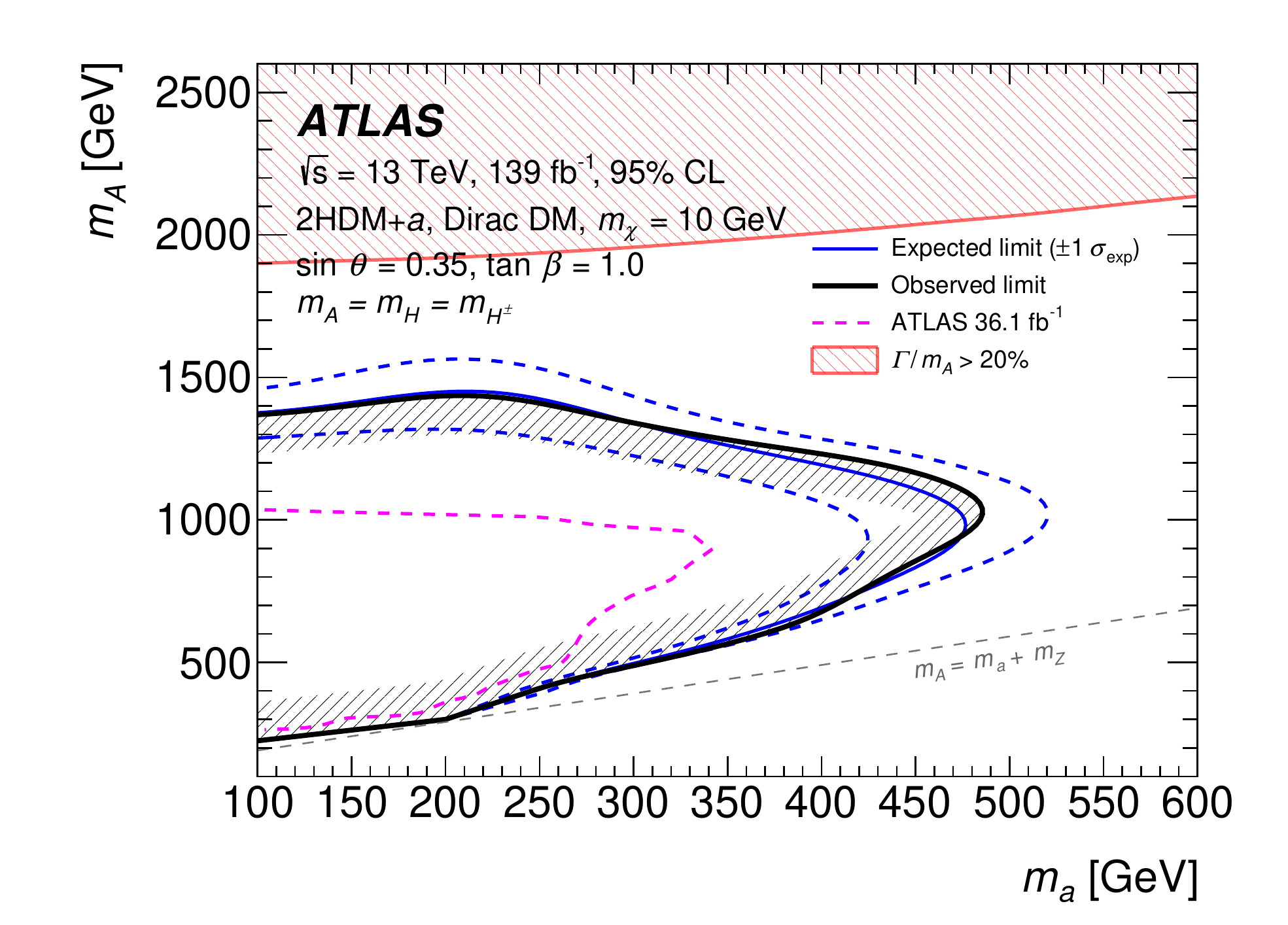} &
\includegraphics[width= 0.45\textwidth]{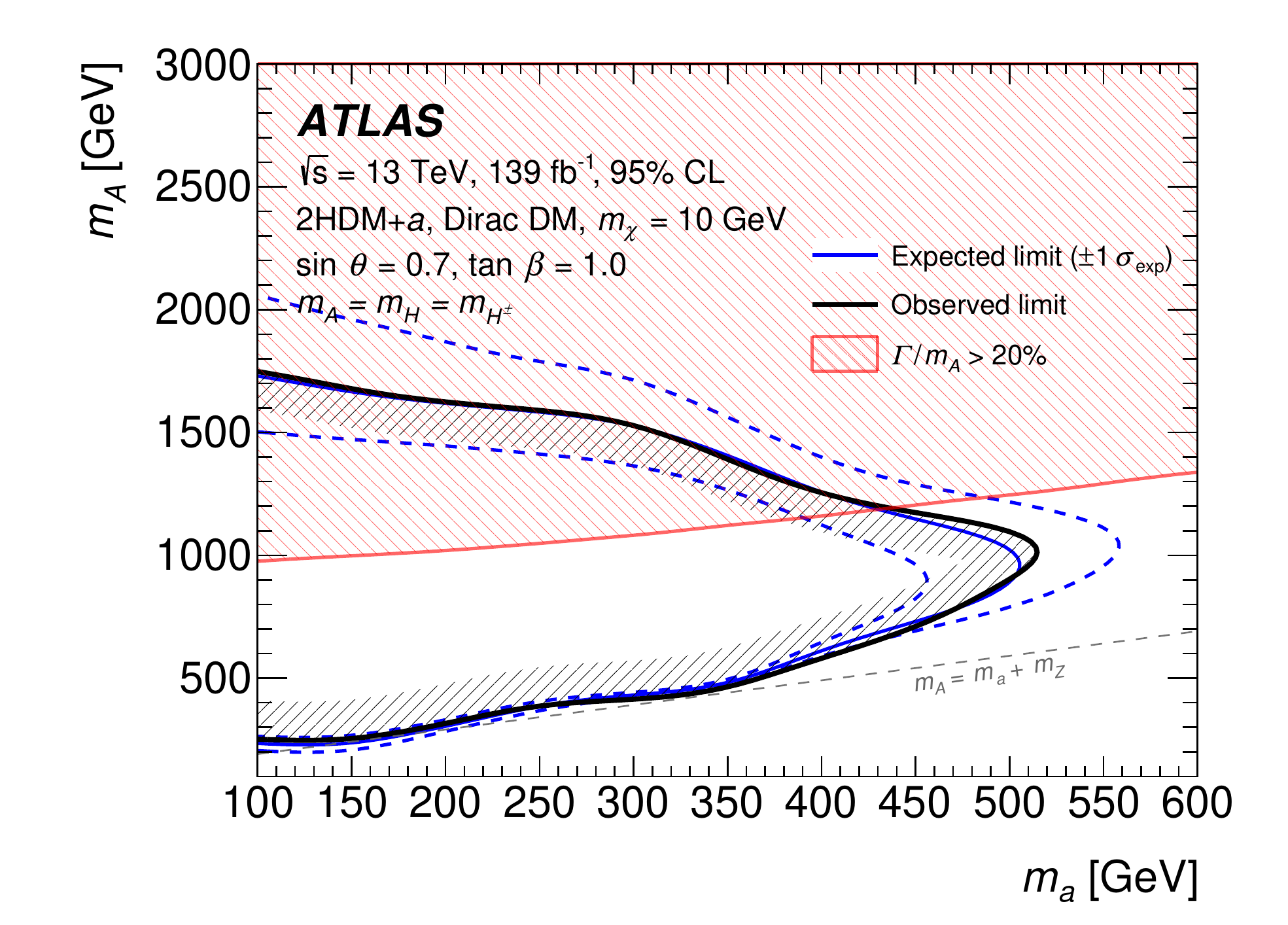} \\
(e) & (f) \\
\end{tabular}
\end{center}
\caption{
Exclusion limits within the context of \HDMa models with different parameter choices and in various planes~\cite{Abe:2018bpo}. Subfigures (a, b) show the \tanBeta vs \ma exclusion limits with $\mA = 600$~\GeV,
(c, d) the \tanBeta vs \mA exclusion limits with $\ma = 250$~\GeV, and
(e, f) the \mA vs \ma exclusion limit with $\tanBeta = 1.0$. The region contained by the solid black line and the hashed black contour is excluded at the 95\% CL. The solid blue line indicates the expected limit in the absence of signal, and the dashed blue lines the corresponding $\pm 1 \sigma$ uncertainty band. Subfigures (a, c, e) assume \hbox{$\sinTheta = 0.35$}, while (b, d, f) assume \hbox{$\sinTheta = 0.7$}. Where included, the dashed magenta lines represent the 36.1~\invfb results from Ref.~\cite{EXOT-2017-32}. The hashed red area indicates that the width of one of the Higgs bosons is larger than 20\% of its mass.
}
\label{fig:2HDMaLimits}
\end{figure}

\begin{figure}[!htb]
\begin{center}
\begin{tabular}{cc}
\includegraphics[width= 0.45\textwidth]{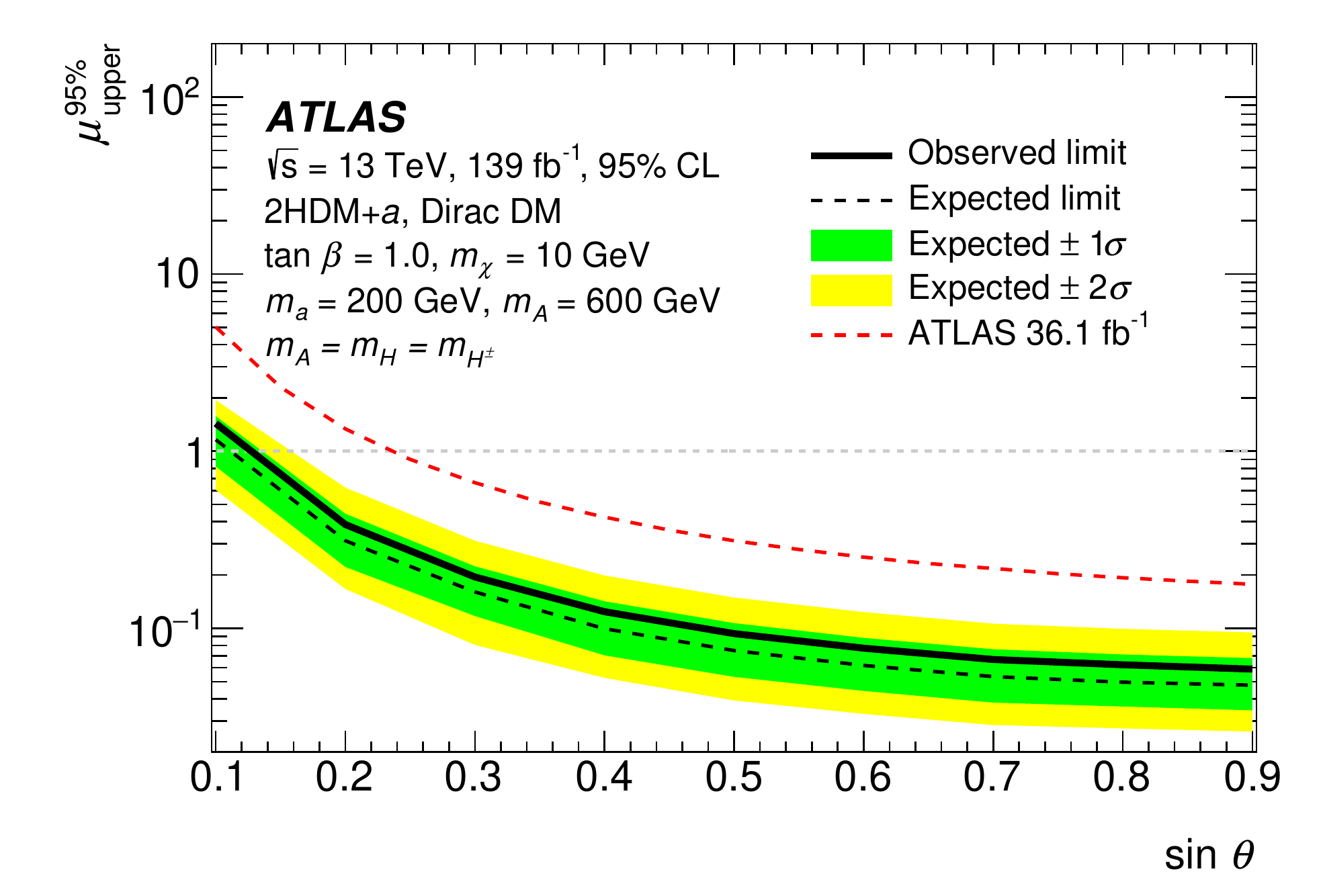} &
\includegraphics[width= 0.45\textwidth]{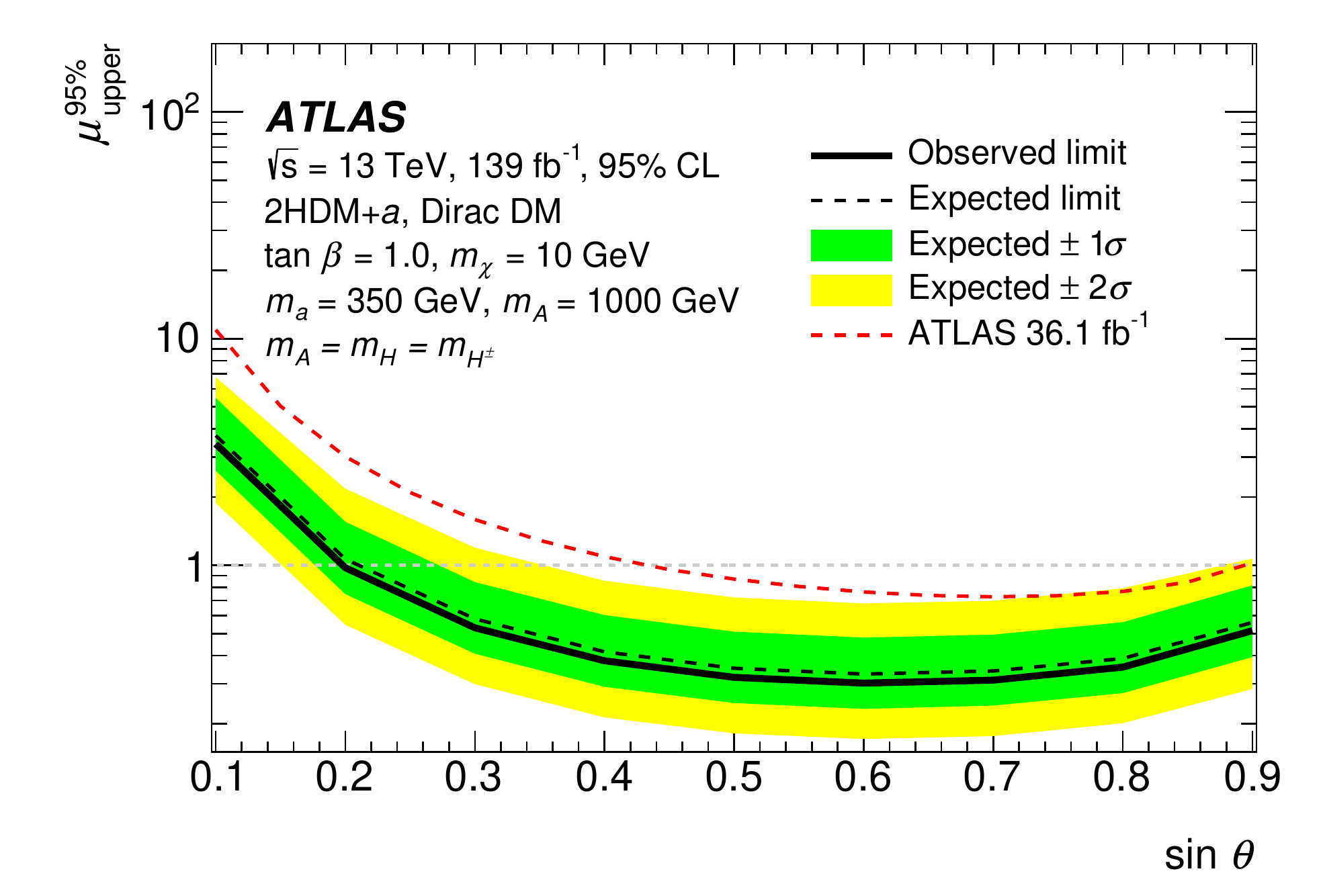} \\
(a) & (b) \\
\end{tabular}
\end{center}
\caption{\sinTheta exclusion limits for \HDMa signals with $\tanBeta = 1.0$ and $\mchi = 10$~\GeV. The solid black line shows the observed limit, while the dashed black line indicates the expected limit in the absence of signal, with the corresponding 1$\sigma$ and 2$\sigma$ uncertainty bands in green and yellow. The dashed red line shows the result from the 36~\invfb analysis~\cite{EXOT-2017-32}. The region below $\mu^{95\%}_\text{upper} = 1$ is excluded at the 95\% CL. In (a), $\mA = 600$~\GeV\ and $\ma = 200$~\GeV, while in (b) $\mA = 1000$~\GeV\ and $\ma = 350$~\GeV.}
\label{fig:2HDMaLimits_CLs}
\end{figure}

Eight scans are produced for the \HDMa models, following the recommendations in Ref.~\cite{Abe:2018bpo}, including the requirement $\mA=\mH=\mHpm$. They are shown in Figures~\ref{fig:2HDMaLimits} and \ref{fig:2HDMaLimits_CLs}. The hashed red area indicates that the width of one of the Higgs bosons is larger than 20\% of its mass~\cite{Bauer:2017ota}. The experimental exclusion in those areas is subject to additional theoretical uncertainties, as the dependence of the width on the virtuality of the additional Higgs bosons could significantly alter the inclusive production cross-sections (one of the limitations of the models). For all presented scans, the exclusion limit is stronger for a mixing parameter value of \hbox{\sinTheta = 0.7} because the cross-sections are larger than for \hbox{\sinTheta = 0.35}.
 
Figures~\ref{fig:2HDMaLimits}(a) and~\ref{fig:2HDMaLimits}(b) show \tanBeta vs \ma limit contours with $\mA = 600$~\GeV, while Figures~\ref{fig:2HDMaLimits}(c) and~\ref{fig:2HDMaLimits}(d) show \tanBeta vs \mA contours for $\ma = 250$~\GeV. Compared to previous limits shown by the dashed magenta line~\cite{EXOT-2017-32} in Figure~\ref{fig:2HDMaLimits}(a), the contours now extend upwards beyond $\tanBeta = 3$ due to the inclusion of $bb$-induced signal contributions (see Figure~\ref{fig:feynman}(d) for an example diagram). The interplay between those and the $gg$-fusion processes (Figure~\ref{fig:feynman}(c)) affects the shapes, as the \tanBeta dependence of the coupling of $H$/$A$/$a$ to top quarks (present in $gg$-fusion) differs from that to bottom quarks. The relative difference between \mA and \ma also affects the shape through the $H \rightarrow Z+a$ process.
 
The $(\mA, \ma)$ exclusion limits are shown in Figures~\ref{fig:2HDMaLimits}(e) and~\ref{fig:2HDMaLimits}(f). Compared to the 36.1~\invfb exclusion limit shown by the dashed magenta line~\cite{EXOT-2017-32} in Figure~\ref{fig:2HDMaLimits}(e), the highest excluded \mA has improved from 1~\TeV\ to $\sim$1.4~\TeV\ for $\ma = 100$~\GeV, and the highest excluded \ma has increased from 340~\GeV\ to 480~\GeV\ for $\mA \sim 1.05$~\TeV. Sensitivity is lost when $m_a + m_Z > m_H$, as particles are not produced on-shell anymore.
For the scan in Figure~\ref{fig:2HDMaLimits}(f), points in parameter space evaluated with signal samples that underwent the simulation of the ATLAS detector response were augmented by points obtained through a reweighting of these samples based on generator-level distributions of key observables, namely $\pt^Z$ (the transverse momentum of the $Z$ boson), $\pt^{\chi\chi}$ (the transverse momentum of the $\chi\chi$ system), and $\Delta\phi(Z,\chi\chi)$ (the azimuthal angle between the $Z$ boson and $\chi\chi$ system), similar to the procedure used in Ref.~\cite{Aad:2021qks}.
Figure~\ref{fig:2HDMaLimits_CLs} shows upper limits  on the signal strength, $\mu^{95\%}_\text{upper}$, as a function of \sinTheta for models with $(\mA, \ma) = (600, 200)$~\GeV\ and $(1000, 350)$~\GeV. Any \sinTheta value for which $\mu^{95\%}_\text{upper} < 1$ is excluded.\footnote{$\mu^{95\%}_\text{upper}$ can be interpreted as  $\sigma^{95\%}_\text{upper}/\sigma_\text{theory}$, where $\sigma^{95\%}_\text{upper}$ is the upper limit on the cross-section of the signal model, and $\sigma_\text{theory}$ is the nominal cross-section of the signal model.} The dashed red line indicates the limit from the previous analysis~\cite{EXOT-2017-32}.

\begin{figure}[!htbp]
\centering
\includegraphics[width=0.8\textwidth]{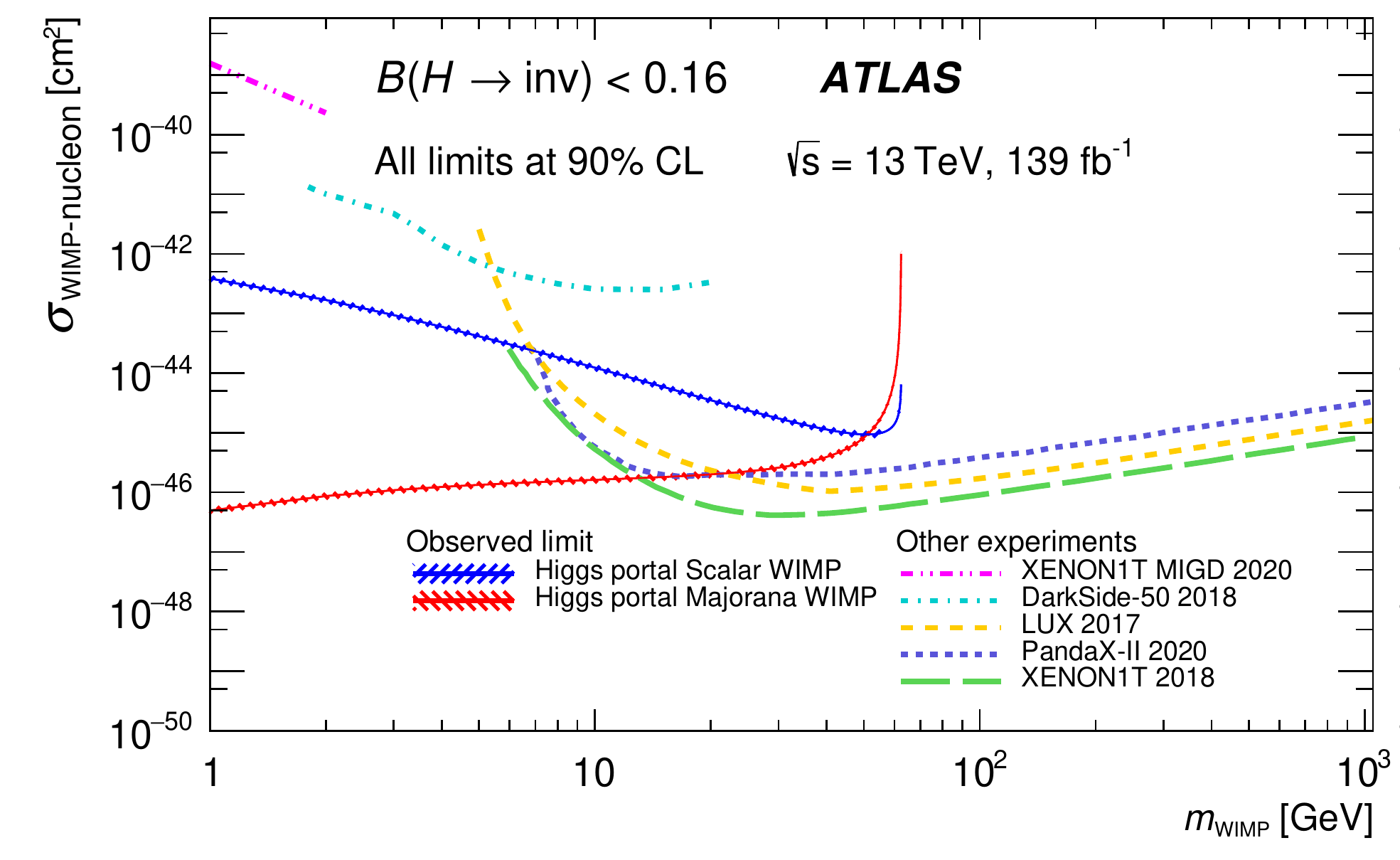}
\caption{\label{fig:ZHinv_direct} Comparison between the 90\% CL upper limits on the spin-independent WIMP--nucleon scattering cross-section from direct-detection experiments~\cite{Akerib:2016vxi,Aprile:2018dbl,Agnes:2018ves,Aprile:2019jmx,PandaX-II:2020oim} and a reinterpretation of the \brHinv limit obtained in this analysis, as a function of the WIMP mass. Higgs portal scenarios are assumed, where the 125~\GeV\ Higgs boson decays into a pair of DM particles~\cite{HIGG-2013-16} that are either scalars or Majorana fermions. The uncertainties from the nuclear form factor are indicated by the hatched band. The regions above the limit contours are excluded in the $\sigma_{\text{WIMP-N}}$ range shown in the plot.}
\end{figure}

\begin{figure}[!htbp]
\centering
\subfloat[][]{\includegraphics[width= 0.49\textwidth]{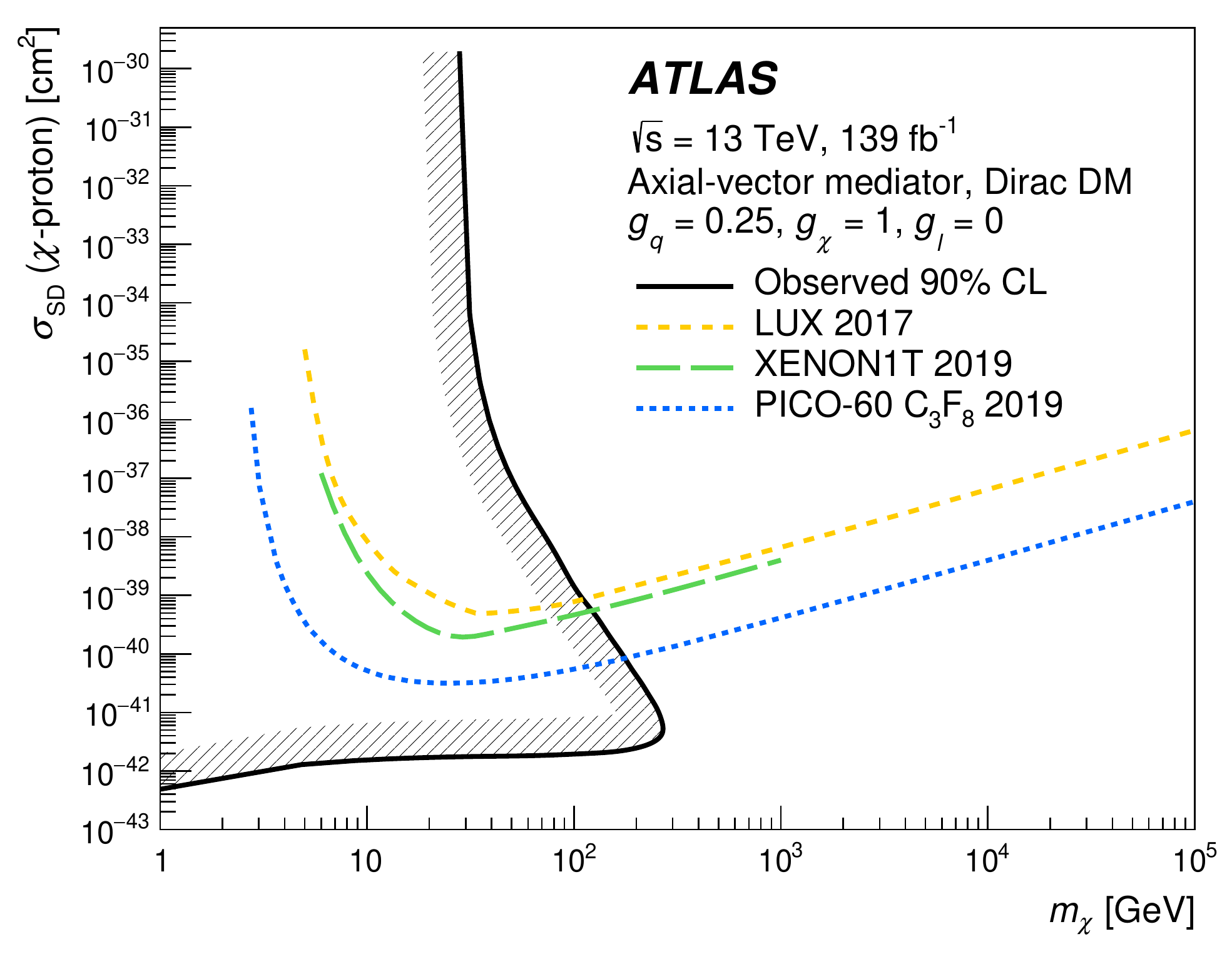}}
\subfloat[][]{\includegraphics[width= 0.49\textwidth]{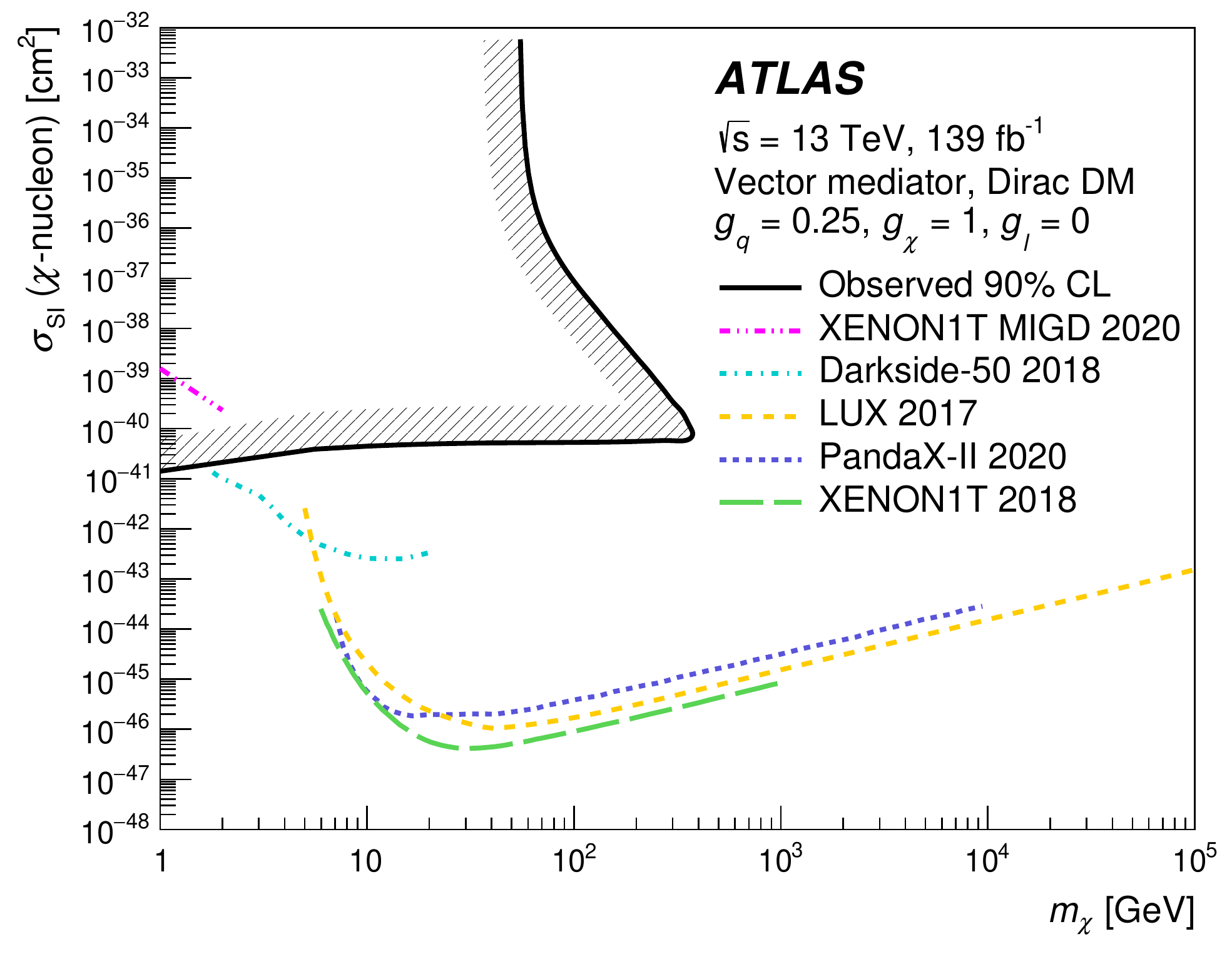}}
\caption{Comparison of the upper limits at 90\% CL from direct-detection experiments~\cite{Akerib:2016vxi,Akerib:2017kat,Agnes:2018ves,Aprile:2018dbl,Amole:2019fdf,Aprile:2019dbj,Aprile:2019jmx,PandaX-II:2020oim} with the exclusion obtained in the simplified DM models in the plane of the dark matter mass and (a) the spin-dependent WIMP--proton scattering cross-section or (b) the spin-independent WIMP--nucleon scattering cross-section~\cite{boveia2016recommendations}. The area within the shaded lines is excluded by this analysis in the context of the simplified models.}
\label{fig:simplified_direct}
\end{figure}

Within the context of appropriate models, both the \brHinv limit and the limits on the simplified DM model (Figure~\ref{fig:simplified_contours}) can be compared with limits from direct-detection DM experiments. For consistent comparisons, the 90\% CL is used, which corresponds to $\brHinv = 16\%$.
The translation of the \Hinv result into a WIMP--nucleon scattering cross-section $\sigma_{\text{WIMP-N}}$ in the Higgs portal model~\cite{HIGG-2013-16} relies on an effective field theory approach and assumes that Higgs boson decays into a pair of DM particles are kinematically possible and that the DM particle is a scalar or a Majorana fermion~\cite{Eboli:2000ze,Fox:2011pm,deSimone:2014pda}. In this translation, the nuclear form factor $f_{N} = 0.308 \pm 0.018$~\cite{Hoferichter:2017olk} is used.
The simplified models with an axial-vector mediator can be translated to spin-dependent WIMP--proton scattering, while the vector mediator induces spin-independent  WIMP--nucleon interactions~\cite{boveia2016recommendations}.
Figures~\ref{fig:ZHinv_direct} and \ref{fig:simplified_direct} show the
complementarity of the collider and direct-detection experiments: the limits obtained in this analysis are particularly competitive for low DM masses, where experiments relying on recoil measurements have limited sensitivity.

% End of text imported from the .//sections/results.tex input file
 
\FloatBarrier
 
\section{Conclusion}
\label{sec:conclusion}
% The next lines are included from the .//sections/conclusions.tex input file
This article presents a search for invisible decays of the Higgs boson as well as searches for dark matter candidates, produced together with a leptonically decaying $Z$ boson.
The analysis was performed using proton$-$proton collisions at a centre-of-mass energy of 13~\TeV, delivered by the LHC, corresponding to an integrated luminosity of 139~\invfb and recorded by the ATLAS experiment. Assuming Standard Model cross-sections for $ZH$ production, the upper limit in the
branching ratio of the Higgs boson to invisible particles was constrained to \Hinvobs, at the 95\% confidence level. The corresponding expected limit of \Hinvexp represents an improvement of about 45\%
in comparison with  a projection of the previous
analysis  scaled to the present integrated luminosity. Exclusion limits were also set for simplified dark matter models and \HDMa models for a number of benchmark parameters, improving on previously set combined constraints obtained with a 36~\invfb dataset.
% End of text imported from the .//sections/conclusions.tex input file

\section*{Acknowledgements}
% The next lines are included from the .//acknowledgements/Acknowledgements.tex input file

We thank CERN for the very successful operation of the LHC, as well as the
support staff from our institutions without whom ATLAS could not be
operated efficiently.
 
We acknowledge the support of
ANPCyT, Argentina;
YerPhI, Armenia;
ARC, Australia;
BMWFW and FWF, Austria;
ANAS, Azerbaijan;
SSTC, Belarus;
CNPq and FAPESP, Brazil;
NSERC, NRC and CFI, Canada;
CERN;
ANID, Chile;
CAS, MOST and NSFC, China;
Minciencias, Colombia;
MEYS CR, Czech Republic;
DNRF and DNSRC, Denmark;
IN2P3-CNRS and CEA-DRF/IRFU, France;
SRNSFG, Georgia;
BMBF, HGF and MPG, Germany;
GSRI, Greece;
RGC and Hong Kong SAR, China;
ISF and Benoziyo Center, Israel;
INFN, Italy;
MEXT and JSPS, Japan;
CNRST, Morocco;
NWO, Netherlands;
RCN, Norway;
MEiN, Poland;
FCT, Portugal;
MNE/IFA, Romania;
JINR;
MES of Russia and NRC KI, Russian Federation;
MESTD, Serbia;
MSSR, Slovakia;
ARRS and MIZ\v{S}, Slovenia;
DSI/NRF, South Africa;
MICINN, Spain;
SRC and Wallenberg Foundation, Sweden;
SERI, SNSF and Cantons of Bern and Geneva, Switzerland;
MOST, Taiwan;
TAEK, Turkey;
STFC, United Kingdom;
DOE and NSF, United States of America.
In addition, individual groups and members have received support from
BCKDF, CANARIE, Compute Canada and CRC, Canada;
COST, ERC, ERDF, Horizon 2020 and Marie Sk{\l}odowska-Curie Actions, European Union;
Investissements d'Avenir Labex, Investissements d'Avenir Idex and ANR, France;
DFG and AvH Foundation, Germany;
Herakleitos, Thales and Aristeia programmes co-financed by EU-ESF and the Greek NSRF, Greece;
BSF-NSF and GIF, Israel;
Norwegian Financial Mechanism 2014-2021, Norway;
NCN and NAWA, Poland;
La Caixa Banking Foundation, CERCA Programme Generalitat de Catalunya and PROMETEO and GenT Programmes Generalitat Valenciana, Spain;
G\"{o}ran Gustafssons Stiftelse, Sweden;
The Royal Society and Leverhulme Trust, United Kingdom.
 
The crucial computing support from all WLCG partners is acknowledged gratefully, in particular from CERN, the ATLAS Tier-1 facilities at TRIUMF (Canada), NDGF (Denmark, Norway, Sweden), CC-IN2P3 (France), KIT/GridKA (Germany), INFN-CNAF (Italy), NL-T1 (Netherlands), PIC (Spain), ASGC (Taiwan), RAL (UK) and BNL (USA), the Tier-2 facilities worldwide and large non-WLCG resource providers. Major contributors of computing resources are listed in Ref.~\cite{ATL-SOFT-PUB-2021-003}.
 
% End of text imported from the .//acknowledgements/Acknowledgements.tex input file

\printbibliography
 
\clearpage % ATLAS Collaboration author list
% Reference date of HIGG-2018-26 is 2021-07-07
% Author list last updated on date 27-APR-22
% Data extracted on 27-Apr-2022 for paper reference HIGG-2018-26
% at 12:05pm
 
\begin{flushleft}
\hypersetup{urlcolor=black}
{\Large The ATLAS Collaboration}

\bigskip

\AtlasOrcid[0000-0002-6665-4934]{G.~Aad}$^\textrm{\scriptsize 98}$,    
\AtlasOrcid[0000-0002-5888-2734]{B.~Abbott}$^\textrm{\scriptsize 124}$,    
\AtlasOrcid[0000-0002-7248-3203]{D.C.~Abbott}$^\textrm{\scriptsize 99}$,    
\AtlasOrcid[0000-0002-2788-3822]{A.~Abed~Abud}$^\textrm{\scriptsize 34}$,    
\AtlasOrcid[0000-0002-1002-1652]{K.~Abeling}$^\textrm{\scriptsize 51}$,    
\AtlasOrcid[0000-0002-2987-4006]{D.K.~Abhayasinghe}$^\textrm{\scriptsize 91}$,    
\AtlasOrcid[0000-0002-8496-9294]{S.H.~Abidi}$^\textrm{\scriptsize 27}$,    
\AtlasOrcid[0000-0002-9987-2292]{A.~Aboulhorma}$^\textrm{\scriptsize 33e}$,    
\AtlasOrcid[0000-0001-5329-6640]{H.~Abramowicz}$^\textrm{\scriptsize 157}$,    
\AtlasOrcid[0000-0002-1599-2896]{H.~Abreu}$^\textrm{\scriptsize 156}$
% ...
\bigskip
\\

% ...
$^{27}$Physics Department, Brookhaven National Laboratory, Upton NY; United States of America.\\
% ...
$^{33}$$^{(a)}$Facult\'e des Sciences Ain Chock, R\'eseau Universitaire de Physique des Hautes Energies - Universit\'e Hassan II, Casablanca;$^{(b)}$Facult\'{e} des Sciences, Universit\'{e} Ibn-Tofail, K\'{e}nitra;$^{(c)}$Facult\'e des Sciences Semlalia, Universit\'e Cadi Ayyad, LPHEA-Marrakech;$^{(d)}$LPMR, Facult\'e des Sciences, Universit\'e Mohamed Premier, Oujda;$^{(e)}$Facult\'e des sciences, Universit\'e Mohammed V, Rabat;$^{(f)}$Mohammed VI Polytechnic University, Ben Guerir; Morocco.\\
$^{34}$CERN, Geneva; Switzerland.\\
% ...
$^{51}$II. Physikalisches Institut, Georg-August-Universit\"{a}t G\"ottingen, G\"ottingen; Germany.\\
% ...
$^{91}$Department of Physics, Royal Holloway University of London, Egham; United Kingdom.\\
% ...
$^{98}$CPPM, Aix-Marseille Universit\'e, CNRS/IN2P3, Marseille; France.\\
$^{99}$Department of Physics, University of Massachusetts, Amherst MA; United States of America.\\
% ...
$^{124}$Homer L. Dodge Department of Physics and Astronomy, University of Oklahoma, Norman OK; United States of America.\\
% ...
$^{156}$Department of Physics, Technion, Israel Institute of Technology, Haifa; Israel.\\
$^{157}$Raymond and Beverly Sackler School of Physics and Astronomy, Tel Aviv University, Tel Aviv; Israel.\\
% ...
$^{a}$ Also at Borough of Manhattan Community College, City University of New York, New York NY; United States of America.\\
$^{b}$ Also at Bruno Kessler Foundation, Trento; Italy.\\
$^{c}$ Also at Center for High Energy Physics, Peking University; China.\\
$^{d}$ Also at Centro Studi e Ricerche Enrico Fermi; Italy.\\
$^{e}$ Also at CERN, Geneva; Switzerland.\\
$^{f}$ Also at D\'epartement de Physique Nucl\'eaire et Corpusculaire, Universit\'e de Gen\`eve, Gen\`eve; Switzerland.\\
% ...

\end{flushleft}

% Created with Glance <Atlas.Glance@cern.ch>

\end{document}